# Lessons Learned in Performing a Trustworthy AI and Fundamental Rights Assessment.

*"Responsible use of AI" Pilot Project with the Province of Fryslân, Rijks ICT Gilde & the Z-Inspection® Initiative.*

Marjolein Boonstra, Frédérick Bruneault, Subrata Chakraborty, Tjitske Faber, Alessio Gallucci, Eleanore Hickman, Gerard Kema, Heejin Kim, Jaap Kooiker, Elisabeth Hildt, Annegret Lamadé, Emilie Wiinblad Mathez, Florian Möslein, Genien Pathuis, Giovanni Sartor, Marijke Steege, Alice Stocco, Willy Tadema, Jarno Tuimala, Isabel van Vledder, Dennis Vetter, Jana Vetter, Magnus Westerlund, Roberto V. Zicari.

*"The results of this pilot are of great importance for the entire Dutch government, because we have developed a best practice with which administrators can really get started, and actually incorporate ethical values into the algorithms used."*

— Rijks ICT Gilde - Ministry of the Interior and Kingdom Relations (BZK)

## Abstract

This report shares the experiences, results and lessons learned in conducting a pilot project "Responsible use of AI " in cooperation with the Province of Friesland, Rijks ICT Gilde- part of the Ministry of the Interior and Kingdom Relations (BZK) (both in The Netherlands) and a group of members of the Z-Inspection® Initiative. The pilot project took place from May 2022 through January 2023. During the pilot, the practical application of a deep learning algorithm from the province of Fryslân was assessed. The AI maps heathland grassland by means of satellite images for monitoring nature reserves. Environmental monitoring is one of the crucial activities carried on by society for several purposes ranging from maintaining standards on drinkable water to quantifying the $CO_2$ emissions of a particular state or region. Using satellite imagery and machine learning to support decisions is becoming an important part of environmental monitoring. The main focus of this report is to share the experiences, results and lessons learned from performing both a Trustworthy AI assessment using the Z-Inspection® process and the EU framework for Trustworthy AI, and combining it with a Fundamental Rights assessment using the Fundamental Rights and Algorithms Impact Assessment (FRAIA) as recommended by the Dutch government for the use of AI algorithms by the Dutch public authorities.

# 1. AI, Ethics, Fundamental Human Rights and the Law of AI.

This section gives a broad introduction and a framing for the use in this report of AI, ethics, fundamental human rights and the law of AI.

## 1.1 The relationship between AI and Human Rights

The question of the relationship between AI and human rights raises the important issue of the relationship between ethics and law. If AI is disruptive in the interpretation and application of the law, the dual aspect of human rights (both positive law and philosophical concepts) allows us to think about the ethical implications of the latter. Insofar as ethics can play different roles in relation to law, it is important to clearly define its different functions, which will at the same time allow us to circumscribe the more precise terrain of the discussion. To do so, the conceptual tools proposed by Luciano Floridi (Floridi, 2018) on digital governance will be used.

This discussion is part of the theoretical framework of information ethics developed by Floridi in his numerous articles and synthesized in his book on the Ethics of Information (Floridi, 2013), which also constitutes a novel contribution to the debate in AI ethics, notably because it addresses some of the shortcomings of classical ethical frameworks (Bruneault & Laflamme, 2021, 2022). Given the limitations of this presentation, we will retain for the purposes of the discussion only a conceptual distinction proposed by Floridi (Floridi, 2018). For the latter, because of the profound transformations that digital technology in general and AI in particular are inducing in our ways of functioning and also because of the many risks associated with them, it is imperative that we develop an adequate normative framework for these technologies, which will undoubtedly occupy an increasing place in the information societies in which the generations that will follow us will evolve. Although these considerations on digital governance are broader in scope than our assessment, i.e. the governance of AI and the Z-Inspection® process (Zicari et al., 2021), they are nonetheless very useful in identifying their components and situating them in relation to each other.

First, for Floridi, digital governance should not be considered as a synonym for digital regulation. In fact, such an adequacy would be a fallacious synecdoche in which the part (regulation) would be unduly substituted for the whole (governance), thus masking an essential part of what Floridi calls „the normative map" of digital governance, namely digital ethics. Indeed, digital governance is not just about making laws and regulations: it must also include the moral evaluation of the issues associated with these technologies, with the aim of proposing specific solutions to the problems under analysis, which also directly links his approach to the pragmatist tradition. According to Floridi, even the presence of appropriate legislative mechanisms would not be sufficient to adequately regulate the development of digital technology and AI, as these mechanisms are limited to determining what is legal and illegal, without questioning the avenues that would be more desirable to follow with regard to technological development. It is of course ethics, drawing on the rich conceptual heritage of moral philosophy, that can assume this function.

Second, for Floridi, the ethics of the digital world can be expressed in two ways in relation to the law: what he calls hard ethics and soft ethics. By hard ethics, he means the discussion on the duties and moral responsibilities of each individual and, more generally, the reflection on the principles and values that should guide moral action. Thought in its relation to law in the governance of digital technology, the role of hard ethics, situated at a higher level of abstraction, is to define the principles that should guide legislative reforms aiming at better framing the conception and deployment of digital technology and AI in society, as well as to question the moral validity of the legislative framework in place. The objective of hard ethics is then to evaluate the coherence of existing laws with the identified ethical principles and to pronounce on their relevance or on the potential need to reform them. In this sense, situated in a way upstream of the law, hard ethics could be likely to influence the legislator's orientations and thus indirectly shape the law. Soft ethics, on the other hand, is situated downstream of the law, i.e. it is interested in ethical questions that go beyond the field covered by the regulations and seeks, for example, to determine, through ethical assessment processes, what technological developments are desirable and what are not, beyond what the law permits or prohibits.

Soft ethics is therefore a practical exercise in the ethical assessment of specific technological devices in concrete situations.
These reflections can also be carried out on the basis of factual or empirical analyses, by means of ethical risk assessment processes in professional environments or decision support tools (Floridi & Strait, 2021). An assessment of ethical risks and impacts conducted in an appropriate manner from the conception of technological devices is likely to guide their development, and consequently to have a real and effective normative effect on practices, for example where the law is silent or absent. Soft ethics, understood as a source of social regulation, is here clearly envisaged in a pragmatic perspective and can be thought of as a source of normativity complementary to law within the framework of normative pluralism. Even if we have to keep in mind that these two functions of ethics cover the same normative ground and that the interactions between soft and hard ethics can be of several kinds, the interest of the distinction proposed by Floridi is that it allows us to identify and situate two distinct normative functions of ethics in relation to law and human rights in AI governance. In this sense, the Z-Inspection® process is a soft ethics assessment, while the use of the Fundamental Rights and Algorithms Impact Assessment (FRAIA) framework (Gerards et al., 2022)in this use case brought a hard ethics aspect to the assessment, more specifically regarding the relationship between AI and human rights as a hard ethics concept (the assessment was not relying on a legal analysis).

## 1.2 The relationship between Ethics and the Law of AI

Ethics and law of AI address the same domain, namely, the present and future impacts of AI on individuals, society, and the environment. Both consider the extent to which AI may enhance or constrain individuals and social initiatives and contribute to or detract from valuable individuals and social interests. Both are meant to provide normative guidance, proposing rules and values on which basis to govern human action and determine the constraints, structures and functions of AI-enabled socio-technical systems. This raises the issue of how to deal with the demands of ethics and law, which may and should indeed converge, but occasionally may pull in indifferent directions.

The law may have failed to adapt to ethical requirements, for instance, not having been able to cope with technological and social development. As a consequence, behavior that should ideally be prohibited (e.g., facial recognition in public spaces) may be legally permissible, or behavior that should be permissible (e.g. processing personal data for the purpose of medical research) may be legally prohibited.

An important connection between morality and law is provided by human rights. Following Sen (Sen, 2004) we may say that human rights are primary ethical demands, in the sense of hard ethics described above. They concern freedoms, broadly understood as opportunities for individuals. Such opportunities include both negative liberties — which mainly require non-interference from state actors and protection from interference by third parties (as in the case of civil and political rights such as freedom of movement and freedom of expression)—and positive liberties, which require the active provision of resources (as in the case of socio-economic rights, for example, rights to education,health, and housing). Human rights also have a legal dimension, i.e., certain important aspects of ethical human rights are also recognised in binding international, regional and national instruments, creating enforceable legal obligations for states and other actors.

Thus a significant overlap exists between (different constructions of) ethical and legal rights, but the two dimensions are distinct. In particular, certain aspects of human rights discourse in ethics may not be explicitly addressed in certain legal systems. This may happen because the law wrongly fails to appropriately adopt ethical standards that it should implement, but it may also happen because the law rightly does not enforce aspects of ethics that are better left to voluntary initiatives inspired by morality or influenced by informal social pressure. As an example of a case in which the law had not, until recently, explicitly endorsed an ethical human right, consider the right to a healthy environment, only recently recognised by the United Nation Right Council. An example of a moral right that the human rights to freedom of expression and information may not be respected in authoritarian legal systems, or the right to education may fail to be implemented in legal systems that do not ensure the universal accessibility of it (even where resources would be available). This distinction between ethics and law does not exclude that the two dimensions may influence each other. Ethico-political arguments can be advanced concerning the need that an ethical right (or aspects or implications of it) should, or should not, be legally recognised, and that the law should change accordingly. Ethical arguments can also be deployed to support the interpretation/construction of legal sources and may thus contribute to determining the way in which the law is applied.

On the other hand, ethics can learn from the law, which takes institutional approaches to normative issues, is expressed in publicly accessible sources and contains vast examples of how (the norms extracted from) such sources are applied to concrete cases. Consider, for instance, how general ideas supporting an ethical right to privacy or an ethical right to free speech and to protection from discrimination have been translated into corresponding legal rights set forth in legislation and upheld in a vast case law.

The continuum between ethics and law is borne out by the fact that when we speak of the impact of AI on broadly scoped rights, such as privacy or freedom of expression, or on collective values, such as democracy, public discourse, public health, or culture, often we do not point to any specific ethical theory or municipal law, we rather refer to a broad cluster of issues, claims, and concepts pertaining to different ethical approaches and different international, regional, or national legal systems.
Multiple references to the rights-language should not be condemned, as it contributes to the richness of the normative debate on the impacts of AI and should be combined with the ability to draw the necessary distinction when needed.

Thus, lawyers should not be worried when the language of rights and values is deployed by ethicists, as when the term 'human rights', or terms such the 'right to autonomy', the 'right to privacy', or 'dignity' appear in documents on the ethics of AI. However, lawyers should refrain from translating ethical claims directly into legal claims. Ethical claims should not be misconstrued as legal claims nor rejected for not being affirmed by existing laws. Similarly, ethicists should not be too impatient when lawyers are slow or reluctant to incorporate, into the law, ethical claims concerning present and prospective uses of AI.

Finally, neither ethics nor law should be viewed as functionally equivalent, namely, as interchangeable substitutes in the regulation of AI. It has indeed been observed that the enthusiasm of the major commercial players for ethical charters may be motivated by the purpose of preventing the enactment of binding laws governing their activity, and consequent institutional controls.

The law is needed whenever only a coercible public response can effectively counter abuses and misuses of AI, as well as when the allocation of public funds, and the deployment of governmental resources has to be directed to support the creation and accessibility of valuable technological solutions. Thus, the adoption of ethical guidelines by private actors does not exempt them from being subject to old and new legal constraints. Similarly, even under an adequate legal regulation of AI, still it makes sense to develop ethical frameworks, to guide the legally permissible uses of AI toward socially beneficial outcomes, and to support the application and evolution of the law.

A fundamental step in legal regulation of AI is now becoming reality in the EU, where the process for the enactment of a regulation on AI, the AI Act is being completed, as the proposal prepared by the Commission has been examined by the EU Parliament and Council, who have reached a common position (Council of the European Union & European Council, 2023; European Commission, 2021). The AI Act adopts a risk-based approach, i.e., aims at ensuring that AI systems entail an acceptable level of risks: risks (expected hazard) must be minimized and in any case they must be outweighed by opportunities (expected benefits). It classifies AI systems into three categories of risk depending on whether they entail: (1) unacceptable risk; (2) high risk; (3) low risk (4) minimal risk. Systems entailing unacceptable risks are prohibited. For instance, this is the case for biometric recognition in public spaces (though with some exceptions). Systems entailing low risks are only subject to requirements of transparency, i.e., that humans are informed that they are interacting with AI systems.

The core of the AI Act pertains to the high risk systems, which are subject to the assessment of the acceptability of the risks they entail. This assessment is to be performed primarily by developers (with some involvement of deployers), and is to be reviewed by certifications bodies. The process is being supervised by national AI authorities, to be coordinated by an EU authority.

The AI Act specifies that the risks to be considered are not limited, as it is usual in market regulation, to hazards concerning health and safety. On the contrary they also include the " fundamental rights, democracy, the rule of law and the environment." An amendment proposed by the EU Parliament explicitly requires deployers of high-risk system to engage in a "human rights assessment" which includes identifying "the reasonably foreseeable impact on fundamental rights of putting the high-risk AI system into use" and "a detailed plan as to how the harms and the negative impact on fundamental rights identified will be mitigated".

Thus, the AI Act sets a formidable task for developers, deployers and certifiers of AI system: identifying and possibly quantifying the risks that their systems affect human rights and social values and minimizing such risks through appropriate mitigation measures, ensuing that expected benefits outweigh any residual risks, while taking into account requirements of usability and economic sustainability.

# 2. Performing a Trustworthy AI and Fundamental Rights Assessment: A Pilot Project

Pilot Project "Responsible use of AI" Pilot Project with the Province of Fryslân, Rijks ICT Gilde & the Z-Inspection® Initiative.

## 2.1 Reason for the pilot

During the conference "AI and the future of Europe" held in Brussels on March 30, 2022, Dutch Secretary of State Alexandra van Huffelen mentioned that the digital transition and the use of AI should always be human-centered and based on democratic values and rights. Governments should lead by example in this regard (van Huffelen, 2022).

The Dutch government wants to seize the opportunities of AI, but the technology still raises many important questions. How reliable are algorithms? Can an algorithm discriminate? What are the ethical and social effects of AI and how transparent is its use? In addition, the use of AI must always be human-centered and based on democratic values and rights.

With that comes an impressive number of rules, frameworks and guidelines in the field of AI. How do you apply them in practice? What do you need to pay attention to? And how do you integrate them into the development and use of AI?

The Pilot Project: "Assessment for Responsible Artificial Intelligence" was conducted by a team of experts belonging to the Z-Inspection® initiative together with Rijks ICT Gilde part of the Ministry of the Interior and Kingdom Relations (BZK) and the province of Fryslân (The Netherlands).

During this pilot, the practical application of a deep learning algorithm from the province of Fryslân was assessed. The AI algorithm processes satellite images producing reports, such as segmentation maps, of nature and farm areas for monitoring nature reserves.

The pilot project took place from May 2022 through January 2023. In this pilot a Trustworthy AI assessment was conducted using the Z-Inspection® process (Zicari et al., 2021), combined with the Fundamental Rights and Algorithms Impact Assessment (FRAIA) (Gerards et al., 2022). FRAIA is recommended by the Dutch government for the Fundamental Rights assessment part.

The "Assessment for Trustworthy AI" pilot sought to answer to the following questions:
1. As a government, how do you govern the development and use of responsible AI?
2. What frameworks, laws and regulations are important, and how do we assess them in the development and use of AI?
3. How do you analyze, assess and improve AI applications?
4. And are the applications consistent with public values and human rights?
5. What ethical issues does the AI system raise?
6. What fundamental rights could be affected by the AI system?
7. What measures could be met for the AI system to be trustworthy?

The pilot gave some answers to these questions and in addition helped to stimulate awareness and dialogue about AI within the Dutch government, and provided guidelines to be able to confidently deploy AI technology for the questions of tomorrow.

## 2.2 The background

The Province of Fryslân is investing, in the coming years, in the smart and effective use of data. The province sees that almost all provincial developments and social tasks contain a data component. This creates urgency in the subject. To be able to responsibly respond to technological developments as a province, a sharp vision on data and AI is needed. Participation in the pilot helped design the future digital infrastructure and outline ethical frameworks.

As directly related to this pilot, the Province of Fryslân is required by law to monitor biodiversity in natural areas. This is done by conducting a manual, visual inspection once every 10 years. There is a need to monitor and map the natural areas more often and monitor heather fields for grassification of heathlands and faster. To facilitate the process, reduce its costs and streamline the procedure, the Province commissioned a third party to develop an AI system for this purpose.

The scope of the pilot was to assess whether the use of this AI system is trustworthy, which fundamental human rights are affected by the AI system, and how it can be used responsibly in practice.

## 2.3 Aim of the AI System

The aim of the AI system is to help ecologists to quickly and frequently image the natural area so that it can be checked whether the intended nature quality objectives are being met, the right management measures can be taken and whether the approach to increasing biodiversity is working.

Specifically the AI system aims to provide information about the diffusion of the invasive and unwanted grass species Molinia caerulea, known as moor grass or pipestraw, and Avenella flexuosa (common name wavy-hair grass) in heather fields using satellite images. The satellite images are made available by The Netherlands Space Office (NSO) on the free Satellite Data Portal where generic high resolution optical satellite images are available to be used in GIS.

The AI system uses descriptive analysis of images which will be available to the unit Nature Information and Nature Management of BIJ12. The prevalence of these grasses in patches on the heather fields are taken as indication of the nitrogen levels, as heather fields are by definition nitrogen-sensitive, and as a consequence of nitrogen deposition the heather field vegetation is quickly overgrown by grasses. The BIJ12 coordinates the national system for monitoring, data storage and information, analysis and reports on nature data and brings information on nature monitoring together for the national government, provinces and nature management organizations involved in nature policy and management. As such, the BIJ12 supports provinces in the execution of legal tasks and with knowledge, information and data about the rural area and the physical environment.

The trustworthy and fundamental rights assessment is done with the understanding that the model will be used for monitoring changes in the heather fields, so that the information obtained from the AI system will help inform:
- nature management plans;
- mandatory EU reporting;
- whether the envisaged nature quality objectives are being realized;
- whether the approach to the nitrogen problem is working; and
- the agreed management measures have had the intended effect.

It is expected that by using a descriptive AI model to classify images using an algorithm that can detect invasive and unwanted pipestraw, the monitoring of nature will be faster, more cost effective and more accurate. However, there are some outstanding questions on whether the information may be used to:
(1) draw conclusions about the effectiveness of particular measures aimed at reducing nitrogen; or
(2) how this AI system might be used in the future, including whether the AI system may be used to inform administrative decisions that could potentially influence how individuals are treated.

During the assessment it was found that special considerations are needed if the model will be used for decisions directly impacting how individuals are treated, for instance the issuance of licenses for economic activities.

It was understood that the information will not be used for establishing attribution (origin of nitrogen). The fundamental rights assessment is done with these boundaries in mind and does not consider if the model is, in the future, used for predictive analysis or used in situations other than what it was developed and trained for.

As a side note. Even though it wasn't the case for this case study, the question regarding "how information generated by an AI system may be used to inform administrative decisions that could potentially influence how individuals are treated" should be taken into account for future projects and future assessment.

## 2.4 Approach

In the pilot, the AI system was examined from three different perspectives:
1. Technical
2. Ecological
3. Ethical and Fundamental Human Rights

The Trustworthy AI assessment was conducted using the Z-Inspection® process (Zicari et al., 2021), combined with the Fundamental Rights and Algorithms Impact Assessment (Gerards et al., 2022).

The project was organized as follows (Fig. 1):
- The Kick off meeting,
- Training of all members of the Pilot,
- Definition of the Scope and Boundaries of the Pilot,
- Creation of three working groups,
- Parallel evaluations and common workshops,
- Dialogue and Presentation,
- Final report with recommendations.

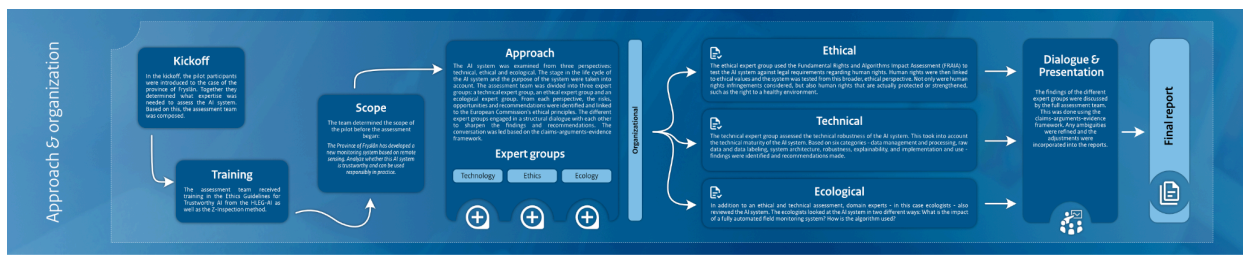

**Figure 1.** *Project flow*

In the kickoff, the pilot participants were introduced to the case of the province of Fryslân. Together they determined what expertise was needed to assess the AI system. Based on this, the assessment team was composed. The assessment team received training in the Ethics Guidelines for Trustworthy AI (AI HLEG, 2019) from the AI HLEG as well as the Z-Inspection® method, and were introduced to the FRAIA.

The team determined the scope of the pilot before the assessment began:

The Province of Fryslân has developed a new monitoring system based on remote sensing. Analyze whether this AI system is trustworthy and can be used responsibly in practice. The AI system was examined from three perspectives: technical, ethical and ecological. The stage in the life cycle of the AI system and the purpose of the system were taken into account. The assessment team was divided into three expert groups: a technical expert group, an ethical expert group and an ecological expert group. From each perspective, the risks, opportunities and recommendations were identified and linked to the European Commission's ethical principles.

The approach used was to provide evidence (whenever possible) for each verifiable claim about the AI (Bloomfield & Netkachova, 2014; Brundage et al., 2020), not limited to the technical part of the assessment, but also to the ethical and fundamental rights part, and for the ecological part.

The three expert groups engaged in a structural dialogue with each other to sharpen the findings and recommendations, in particular when evidence from the technical assessment was relevant for the ethical and fundamental rights assessment. For example, by identifying the resolution of the images, we understood that each single pixel was covering a minimum of 10 meters, effectively removing the presence of any identifying feature and concerns about privacy. Regular common meetings ensured that the partial results of each group were shared with the other groups. From each perspective, the risks, opportunities and recommendations were identified and mapped to the European Commission's ethical principles (AI HLEG, 2019).

From the ecological perspective, the interactions between the Province experts and the Z-Inspection® ecologists team were crucial to conduct a comprehensive review of scientific works and gray literature relevant to the grassification processes in the Netherlands. Baseline information, recent trends, and geographical data pertaining to heather fields were compiled. Meetings were also held with ecologists who had previously conducted surveys for monitoring the grassification processes. During these meetings, common challenges were identified, and the monitoring protocol, as well as ecosystem function characteristics crucial to ecosystem functioning, were discussed.

In a subsequent phase, regular online meetings were held involving ecologists and remote sensing technicians. The classification maps generated using AI algorithms were assessed for accuracy, providing valuable data for evaluating algorithm performance. These discussions allowed for the clarification of pertinent issues and the refinement of the mission and objectives associated with this novel automatic classification approach. Under the ethical perspective, within the Dutch government, the Fundamental Rights and Algorithms Impact Assessment (FRAIA) is an important tool to identify infringements of human rights in the deployment of AI systems. To facilitate an overview of the issues at stake, FRAIA distinguishes between four main areas of concern, or, phrased differently, four relevant clusters of rights (Ulrich, 2023).

These are:
1. Fundamental rights relating to the person (including a number of social and economic fundamental rights)
2. Freedom-related fundamental rights
3. Equality rights
4. Procedural fundamental rights

Each primary cluster is further associated with more specific sub-clusters of rights, which are detailed in an annex to the FRAIA framework, and the aim of an impact assessment is to determine specifically 'which sub-clusters an algorithm affects or may affect.'

The Z-Inspection® process provides a holistic and dynamic framework to evaluate the trustworthiness of specific AI systems at different stages of the AI lifecycle, including intended use, design, and development. It focuses, in particular, on the discussion and identification of ethical issues and tensions through the analysis of *socio-technical scenarios* and a requirement-based framework for ethical and trustworthy AI. Z-Inspection® can be used to co-design, self-assess, or conduct independent audits of AI systems together with the stakeholders owning the use-case (Vetter et al., 2023; Zicari et al., 2022).

The process comprises three phases: (1) set-up, (2) assess, and (3) resolve. A schematic description of them is presented in temporal order in Fig. 2.

The set-up phase consists of the validation of several pre-conditions before the assessment starts including the legal admissibility and absence of conflict of interest, the setup of an interdisciplinary team of experts working together with the key stakeholders owning the specific AI use-case, and finally, the definition of the boundaries and context where the assessment takes place. The assess phase is an iterative process that includes the creation and analysis of socio-technical scenarios, the identification of ethical issues and tensions, the validation of claims by providing evidence (if any), and the mapping to the EU trustworthy AI framework using a mapping from "open to closed vocabulary" as a consensus-based approach. The resolve phase addresses the ethical tensions identified during the assess phase, here possible trade-off solutions are proposed, possible risks and remedies are identified, and recommendations are made to the key stakeholders. A detailed description of the three phases can be found in (Vetter et al., 2023; Zicari et al., 2021).

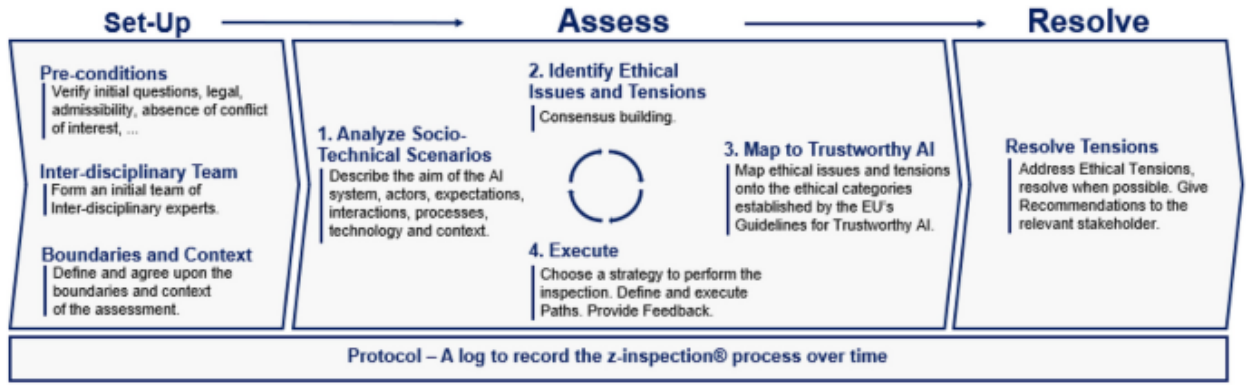

**Figure 2.** *Z-Inspection® process flow chart describing the main steps of the set-up, assess, and resolve phases. In parallel to the phases, a log is kept in which the process and events of the assessment are tracked. Adapted from (Vetter et al., 2023).*

Using the Z-Inspection® process key stakeholders were able to discuss critical issues such as: the purpose of the algorithm, the development process, ethical dilemmas and trade off and possible conflicts of interest. In general, this enables organizations to develop and use responsible AI applications in a structured and accountable manner. Furthermore, it is very important to share the knowledge and experiences from the pilot. First, to stimulate digital awareness and dialogue about AI within the government. And second, to be able to confidently apply the technology to tomorrow's questions.

2.4.1 Technical Assessment

This part of the assessment considered the technical perspectives and issues detected in the system that can potentially give rise to ethical or legal issues and limit the applicability of the system. For the analysis, we considered the provided technical prototype, and were given access to a repository with code and data. During the assessment the maturity of the prototype was defined, by the owner of the project and agreed by the technical group, as technical readiness level six (TRL6) meaning that the prototype as such is not yet ready and prepared for release into a production environment (European Commission, 2014).

For this use case, the technical group assesses the modeling solution's setup and ability to classify the degree of presence of specific species of grass in the input satellite images. The geographical areas used as input data has been selected by experts and is manually verified to a degree, but the labeling process is defined through automated means. Labeling in this case means the categorization of selected pixels as (unwanted) nitrogen-influenced grass or not. We have divided the identified technical issues into the broad categories of data processing, data labeling, system and architecture, robustness, explainability, and accountability. These follow the pillars of ethical AI given by the EU high-level expert group. The identified technical issues should be considered as further studies that are needed to advance the solution to a TRL7 level.

### 2.4.2. Ecological Assessment

Ecological data is traditionally collected through labor-intensive fieldwork. Manual work makes it expensive and impractical for large-scale studies at a fine-grained temporal level. Remote sensing may benefit ecological fieldwork by enabling data acquisition at a large spatial scale and improved temporal resolutions. This pilot project assessed the benefits and potential problems of a pilot monitoring system in a Natura2000 site in the Province of Friesland, in supporting ecological-driven decisions (some of them with potential legal repercussions). The new monitoring system relies on remote sensing and an Artificial Intelligence (AI) algorithm and aims to track the changes in the cover of (unwanted) nitrogen-sensitive grasses (Molinia caerulea and Avenella flexuosa) used as bio-indicators of nitrogen pollution in the protected area. The monitoring system responds to specific ecological demands that are exclusive of heather vegetation. When dwelling on accuracy assessment (benefits and barriers) as part of the evaluation process, it is relevant to keep in mind that satellite-based remote sensing, namely optical data and reflectance responses of the vegetation, and not necessarily the AI per se, is behind some of the differences observed between both systems (fully human-controlled vs. an automated satellite-based system). Attribution of AI performance per se would be better informed if an automated non-AI algorithm was used on the same satellite data. This would allow us to better assess the performance of the AI algorithm per se.

For the above reasons, when assessing the performance of the AI system in this report, we are inevitably looking into two different new methods: 1. the transitioning between a fully manual field-based land monitoring system towards an automated satellite remote sensing-based system. And 2. the use of an AI algorithm that relies on this satellite remote sensing system to produce final maps for the 2 selected species (Molinia caerulea and Avenella flexuosa). In other words, in this evaluation, when we assess the benefits/challenges/barriers of the new AI system, we are combining the use of an AI algorithm + the transition towards a higher automatization in the monitoring system that relies on satellite remote sensing. If the satellite data introduced (e.g. radar, hyperspatial), or the AI algorithm selected is changed or modified, the AI system's performance would change.

### 2.4.3. Ethics and Fundamental Rights Assessment

In light of the introduction of a fundamental rights impact assessment tool for algorithms in the Netherlands in March 2022, a hybrid approach was adopted in the pilot. First, the AI system was assessed against the human rights requirements using the

FRAIA. This assessment did not only consider human rights violations but also rights which could be protected or strengthened by applying the AI system, such as the right to a healthy environment. Then ethical issues were identified and assessed based on the European guidelines for responsible AI and the system was assessed from this broader perspective.

The FRAIA comprises four parts of which the fourth part is on fundamental rights assessment, with a roadmap comprising seven steps, aimed at identifying rights which may be affected by the AI system and any relevant legislation, and assess the seriousness of any potential infringement in relation to the objectives of the AI system.

It was found that there were both overlaps and complementary steps between the Z-Inspection® and the FRAIA. To avoid duplications with other parts of the Z-Inspection® only the questions in step 1 and 3 of the FRAIA roadmap for the fundamental rights assessment were included in the pilot, while the other questions of the fundamental rights assessment were considered in the broader context of the Z-Inspection.

While the FRAIA focuses on fundamental rights, the envisaged approach is, like the Z-Inspection® process aimed at facilitating interdisciplinary dialogue and discussion, and be a decision-making tool for government organizations who commission the development and/or use of an algorithmic system. We therefore found that there was merit in combining the two approaches.

In identifying the fundamental rights being affected by the AI system the assessment looked at the list of fundamental rights provided in the FRAIA, which are clusters around four groups with specific rights listed under each of the areas. Rights related to:
1. The person
2. freedom-related fundamental rights
3. equality rights
4. procedural fundamental rights

Following this step, the assessment considered more broadly ethical issues arising from the AI system. Specifically, the ethics guidelines for trustworthy artificial intelligence were considered as defined by the EU High-Level Expert Group on AI (AI HLEG, 2019). The four ethical principles of the guidance were used, acknowledging that tensions may arise between them:

(1) Respect for human autonomy (2) Prevention of harm (3) Fairness (4) Explicability

Furthermore, the seven requirements of Trustworthy AI defined in (AI HLEG, 2019) were considered. Each requirement has a number of sub-requirements as indicated in Table 1.

**Table 1.** Requirements and sub-requirements Trustworthy AI. Reproduced from (AI HLEG, 2019).

| Requirements Sub-Requirements |
|---|
| 1 Human agency and oversight *Including fundamental rights, human agency and human oversight* |
| 2 Technical robustness and safety *Including resilience to attack and security, fall back plan and general safety, accuracy, reliability and reproducibility* |
| 3 Privacy and data governance *Including respect for privacy, quality and integrity of data, and access to data* |
| 4 Transparency *Including traceability, explainability and communication* |
| 5 Diversity, non-discrimination and fairness *Including the avoidance of unfair bias, accessibility and universal design, and stakeholder participation* |
| 6 Societal and environmental wellbeing *Including sustainability and environmental friendliness, social impact, society and democracy* |
| 7 Accountability *Including auditability, minimization and reporting of negative impact, trade-offs and redress.* |

# 3. Technical and Ecological Assessment

A report has been prepared for each working group. In addition to substantive reports, lessons learned from applying the Z-Inspection® process and FRAIA were also identified. This is summarized in the lessons learned overview below.

## 3.1 The Main Lessons Learned

The following is a description of the main nine lessons learned when conducting this pilot project (Rijksorganisatie voor Ontwikkeling, Digitalisering en Innovatie, 2023).

**1. Clearly define the scope.** It is important to establish a clear scope for the assessment in advance. What will the team assess? But also: what is the team not going to assess? It is tempting to zoom out, to look from a broad perspective and also to evaluate the government policy to which the algorithm relates. During the assessment, therefore, it is critical to ensure a clear scope and a process supervisor who continuously monitors and defines the scope in advance.

**2. Provide a Common Language.** The assessment team is interdisciplinary with different backgrounds and skills. It takes time and patience to understand each other and develop a common language. Mapping findings to the ethical principles of the HLEG-AI framework helps create greater understanding and facilitates dialogue.

**3. Z-Inspection® is more than a method.** Z-Inspection® is more than a method. It is an international community in which there is joint learning on how to put ethics and responsible use of AI into practice.

**4. Z-Inspection® is suitable for high risk Systems.** The Z-Inspection® method is a good addition to the methods and tools already used by the Dutch government. Because of its depth and the knowledge and time required, it is especially suitable for high-risk algorithms.

**5. Increasing digital awareness.** The pilot and the method encourage dialogue about AI within the government, increasing digital awareness among civil servants. The benefit is that civil servants are more likely to understand the impact AI can have on their work, organization and society. It gives them more confidence to navigate the digital world and seize the opportunities of AI without losing sight of the risks. In this way, technology can be used with confidence for tomorrow's questions.

**6. FRAIA and Z-Inspection® strengthen each other.** During the pilot, the assessment team used two different approaches: the Fundamental Rights and Algorithms Impact Assessment (FRAIA) and the Ethics Guidelines for Trustworthy AI. The two go hand in hand.

Both approaches provide critical insights regarding the AI system. Both ethics and human rights are about norms and fundamental values in society. Since ethical reflection and ethical guidelines influence law, experts from both fields must work together when considering the design of AI systems and their societal implications. Ethics, a branch of philosophy, considers what is right and wrong. It seeks answers to questions such as "What should we do?" or "What is the right action? In the context of AI systems, an ethics-based approach focuses on questions such as "What is the right way to design, develop, deploy and use this type of technology so that it benefits individuals and society?

Such questions require reflection on the various courses of action around an AI system, on the different options and their implications. This reflection should not be limited to what legislation prescribes; a broader ethical perspective is needed. A human rights-based approach is closely linked to existing law and focuses primarily on aspects that are legally relevant and enforceable.

**7. Organizations and project teams are looking for strong footing.** Organizations and project teams are looking for strong footing. They often ask for one complete checklist so they can be sure that their AI system is reliable and compliant with all regulations. But do we want the government to follow a checklist and check off all the boxes? No, we want the government to be guided by public values and protect fundamental rights. Therefore, the need for a holistic, integrative and interdisciplinary method for responsible AI is great. Not just a static checklist, but a dynamic process for conducting assessments. An approach that ensures that the assessment team is representative, knowledgeable and independent. A method that facilitates meaningful consultation. A way to assess the reliability of AI systems, supported by arguments and evidence. The Z-Inspection® method meets this need.

**8. AI System validation is essential.** Technical model validation is necessary to gain insight into potential ethical dilemmas. Also, an AI system that is not robust can never be trustworthy. Insight into how the algorithm performs within a real environment and organizational setting is an important prerequisite for responsible use.

**9. Courage.** Working proactively, with an external team of independent experts openly and transparently within government is new, and it causes quite a bit of cold feet. Conducting a self-assessment for responsible AI while providing full disclosure requires courage. Openness helps to learn from each other and steer the deployment of responsible AI in the right direction.

## 3.2 Technical Assessment

This section presents the assessment results of the technical working group. The main focus of the technical evaluation was technical soundness of the complete implementation, including data collection and -management, system architecture and deployment procedures.

### 3.2.1. Training Data

The model was trained on data collected from ESA Sentinel-2 satellite imagery that has an orbital swath width of 290 km. Imagery used is from selected regions of interest chosen in cooperation with ecologists. The imagery spans a time interval from 2016 forward, and the end date is estimated to be around 2020/2021 when the project team acquired the data. The SENTINEL-2 imagery data includes optical data that samples 13 spectral bands: four bands at 10 m, six bands at 20 m and three bands at 60 m spatial resolution. The selected image resolution is 10 m, meaning that lower resolution spectral band data is up-sampled.

For some of the models that the developer team tested with, filtering techniques were used to remove noisy pixels resulting from cloud cover or cloud shadows. Further, fixed-size regular-spaced time series features were used for example to determine season median values per spectral band. However, for the final model no filtering or aggregation was performed. Instead, every pixel is considered a sample, and consecutive images taken over a year are vectorized into a time window. This time window contains pixels spaced in the temporal dimension, utilizing both historic and future information that is available to estimate the categorization at the time of physical inspection.

Figure 3 shows the use of the different data sources, such as manually inspected vegetation maps (kartering 2015), training (Luchtfoto 2016), test data (Luchtfoto 2020), and model output (Voorspelling 2020).

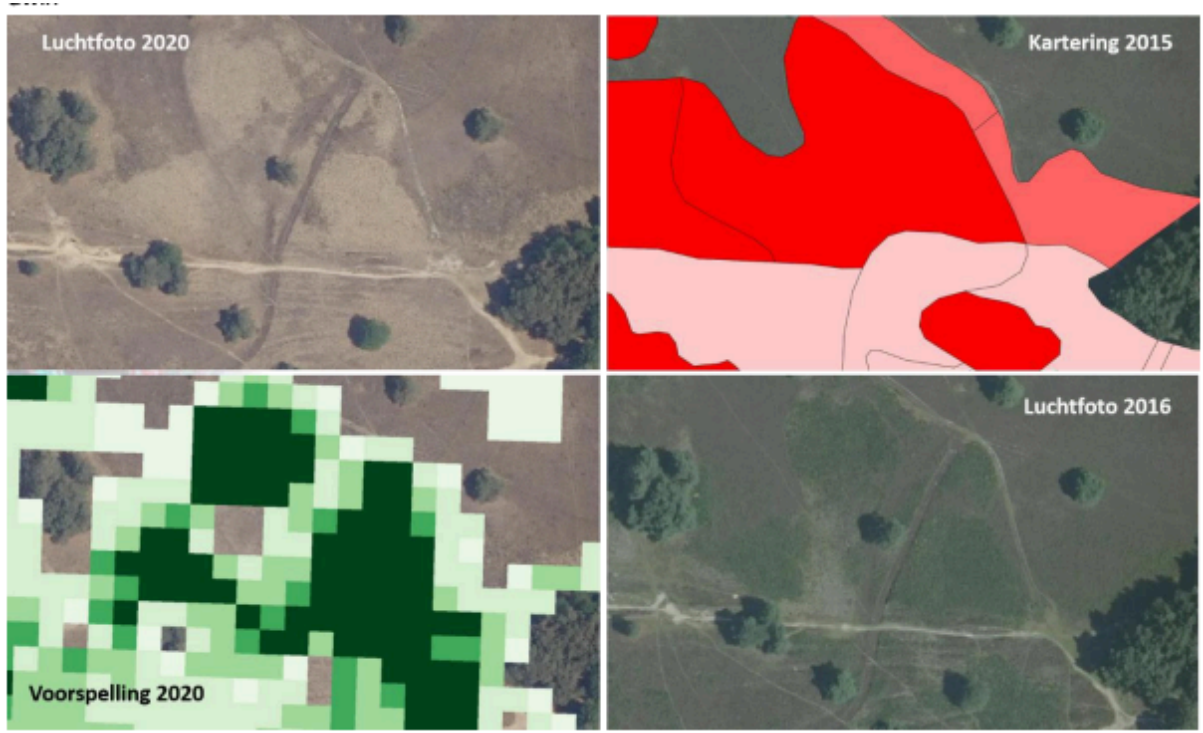

**Figure 3.** *Example satellite images with vegetation maps.*

**Small size of the training data set.** The dataset is used for training an LSTM model (Hochreiter & Schmidhuber, 1997) consisting of two consecutive LSTM cells with 18 and 12 dimensions respectively. The model was trained with a batch size of 128 samples, for 15 epochs, with a learning rate of 0.001. Each epoch consisted of 2000 batches. No regularization was applied. While the training process seems to converge around a sufficiently optimal model solution, the dataset might not be extensive enough to allow for sufficient generalization to areas outside of the chosen areas. A further inspection of good practices defining the number areas needed to train a generalized model may help assess the risks associated with the small number of training areas. While the country is geographically rather homogeneous, a further study is needed to understand the regional variance (e.g., sea shore vs inland) and the implication on model output.

**Data lacks representational fairness.** The dataset was collected in collaboration with ecologists to determine areas of interest. The chosen areas determine a bias in data collection, used to reveal information about targeted areas. This process early on in development allows for expert validation, however, data chosen does not represent the whole country or even the region of Friesland. Selected areas are handpicked for their closeness to moorland, suggesting that a further study of more diverse area types is needed to understand the generalizability of the solution.

*Trade Off:* Annotating and using images from outside current areas, such as borders to the selected areas or new heatlands, might make the model more robust but will also require serious efforts from ecologists.

**Data collection lacks error correction procedures.** Satellite imagery taken from space is sensitive to interference. The final prototype model used seems to work well in terms of noisy data, in fact so well that filtering out noise (e.g., cloud coverage) is not included in the final prototype. A lingering question remains though on how robust this approach will continue to be. A further study is needed to determine the direct influence of noise while retaining accuracy. Robustness and maturity in data collection and data wrangling demands descriptive documentation that details such procedures and demands. Metrics that analyze data quality should be defined that are monitored for both training and testing data. Such metrics could also track data and concept drift over time e.g. for type of vegetation or input satellite images. Such metrics might allow in the future also to sustain a TRL7.

### 3.2.2. Labeling

The labeling effort utilizes previous work by ecologists that have mapped vegetation in certain chosen areas around the Netherlands. The work of defining a ground truth is labor intensive and prone to human errors. Therefore this manual mapping is only conducted once every twelve years. The project tasked with the prototype design is not directly involved with the mapping and therefore, much work has gone into understanding what ecologists have mapped and what type of errors are reoccurring in data. Source data for labeling is selected from vegetation maps from both dry and wet moorlands, spread over NL. The source data includes data from 2015 forward until collection date (assumed until 2020/21).

The sampling choice for source data is based on a selection criterion that aims to find a subset of data that is usable for processing. Only vegetation maps that follow the Staatsbosbeheer (SBB) data schema and that are accessible as a shapefile (geospatial vector data format) with SBB vegetation types are chosen. Further, additional information must be accessible from databases that describe the growth of moor-grass and other relevant vegetation types.

Finally, a manual check per vegetation map is performed that considers the mapping consistency of the source data. Inconsistencies occur and work involves a discussion with ecologists on how to treat such occurrences. Once the manual check is performed, labeling commences by defining the desired outcome.

*Multiclass label.* The first defined outcome is to determine the occurrence of grass in the vegetation maps. This standardizes the different degrees of grassing provided by the ecologists into 6 different bins: 0%, 0-10%, 10-20%, 20-40%, 40-70%, 70-100%. The model generates a probability value between 0 and 1 for all gradations. For each pixel, this is reduced to the most probable gradation. During training an ordinal loss function is used, which means that the degree of difference between the model's prediction and the actual value is considered, i.e. for a true value of 0-10%, a prediction of 20-40% is more accurate than a prediction of 70-100%. During evaluation, an agreement between the model and the mapping of an ecologist is defined as either classifying a region as <50% grassing or >50% grassing.

*Multilabel classification.* The second desired outcome is to understand group types of vegetation. For this purpose, vegetation is categorized as an ordinal multilabel classification, where each bin represents a vegetation shape. Hence, a bin can represent forest shape or other relevant categories.

**Ambiguous ground truth makes validation difficult.** Determining the outcome of a binary classification often demands a clear separation of the classes. In reality, such as in this case, it can often be hard for a human expert to make such a clear separation and further when automating labeling through scripting this is potentially accentuated. Due to the task at hand, determining a label of grass growth depends on numerous other variables that may be correlated directly or indirectly. Considering in particular the temporal events that precedes a manual investigation (e.g., humidity, temperature) may lead to a significant variation in the ground truth labeling of areas that are developing (changing from non-grass to grass or vice-versa). Further, potential temporary and localized weather patterns may influence the outcome of the labeling process. In addition, the human annotators are often performing the labeling for large patches of land at the same time, although the "true" label for some small patches in this area might be different. This potentially leads to inaccurate labels when compared to the much finer resolution used by the AI system and indeed this effect was shown when the AI system's output was validated by ecologists in the field.

**Comparison with labels compared by human experts is lacking.** The model outputs one of 6 values to describe the degree of grassing. However, the original value provided by the ecologist can potentially not cleanly fit into one of the categories the model can output. This makes evaluation difficult. The developers define an agreement between model and ecologist, if either both classify a region as <50% grassing or both classify the region as >50% grassing. A qualitative evaluation with ecologists found that the model almost always correctly detects clearly degraded or non-degraded areas. In addition, coarse errors are limited to a few pixels. For large areas, the 'error' is limited to a difference of one degree (0% is for example seen as 10-20% grassiness). However, the combination of these properties of the model make a reliable estimation of the model's outputs for medium grassing difficult.

**Combining datasets.** In order to link the label data shapes with the satellite imagery, label shapes are rasterized. Once rasterized the label pixels are linked with corresponding satellite pixels. Hence, the input variables to the model are taken from the different bands per pixel over a time window and then the model learns to estimate the label category. The final prototype labels all samples to each available ground-truth.

The train-test split accounts for correlations between neighboring pixels and assigns pixels belonging to the same shape, to the same bin. Pixels are only considered if they are completely inside the vegetation shape, and the rest are discarded. Sampling for each segment aims to assign samples roughly proportional for label / area / year. Hence, segmentation mixes a specific area and year over both train and test segments. We note that spatially the split is very fine-grained in the sense that nearby areas fall into different sets. While we cannot determine the effect of such a split, it might lead to undesired results and metrics that overestimate classifier performance.

**Accuracy does not imply real world performance.** The prototype work shows evidence that the model can detect the degree of presence or absence of specific grass species. However, the unclear labels, ambiguous and border pixels, are as far as we discern outside the scope of the accuracy score. Further metrics are needed to understand these areas, including physical visits for area validation. While we hold it likely that the model is able to identify grass labeled pixels beyond the very clear cases, a ground truth validation must be done to ensure the model's real world performance for tricky situations. Hence, the prototype does not, in its current form, replace the ecologist mapping, but rather complements the ecologist produced mappings.

**Potentially biased.** A common problem of using experts to perform labeling is the introduction of unconscious biases in the labeling process. Ecologists need to receive detailed instructions on how to avoid such pitfalls and control studies can evaluate potential biases by utilizing double or triple mapping, where ecologists are unaware of the others categorization.

### 3.2.3. System and Architecture

The *technical readiness level* (TRL) scale was introduced in the EU for defining project maturity. The scale offers a point of reference for determining the development or maturity of a research and its readiness for the market uptake and potential investments. The levels TRL6 - TRL7 are used while the prototype is in a development stage and transitions towards expected behavior of the future product in a real-world environment. TRL6 is defined as testing the prototype in a realistic and relevant laboratory environment in order to determine that it is representative for the intended operational environment. Hence, TRL6 limits requirements for maintenance processes that are required for higher levels. In our opinion, this adequately describes the current system. The following level, TRL7, aims to reach acceptance for operational use. TRL7 acceptance demands a study of generalizability of the solution towards various test areas that reveal the integrated solution's operational performance. However, such a study is not yet conducted and therefore the AI system has not reached this level yet.

**Monitoring metrics are undefined.** When using deep learning models in production, it is crucial that clear measures are in place for model performance monitoring and procedures for how frequently a model fitness evaluation must be performed.

*Recommendation: Define metrics and actions for performance monitoring.* The system should track a set of metrics that measure how well it performs. Further, a feedback mechanism that allows the solution to be continuously compared against new data from the ecologists is needed. With the help of these metrics, one can define and execute specific actions if metrics drop below a certain threshold, indicating when the system does not perform with the expected accuracy anymore. These metrics can then indicate when a re-training of the system with more recent data is needed.

**No feedback process is in place.** Ecologists have no clear way of notifying the system developers of issues with performance, possible misclassifications, problems with the interface or the explanation mechanism. A dedicated process for giving and including feedback might make it easier for the developers to react to the potential problems and extend their database with new annotated areas. This process could be a simple flagging done by the users/ecologists during regular usage of the system. Those flags are then reviewed and actioned by the technical support/development team. It is also unclear if / how the annotations of updated survey mappings are stored for future model improvements.

**Vocabulary is undefined.** During our discussions we have learned the importance of using a shared AI vocabulary. Initially we spent much of our discussions on understanding what the real claims are and what evidence can support such claims. This issue is common in organizations setting out to introduce AI solutions but can be helped by clearly stating the used terminology in the scoping of the project. A recently released vocabulary summary can assist in finding a shared language (Estévez Almenzar et al. 2022).

### 3.2.4. Robustness

Robustness of the method is unclear. Some more exploration of how the location of the region border influences the scoring might be appropriate and the detected borders should be included as part of the explanation. There was no evidence or comment pointing out if slight changes in the regions could lead to changes in the final scores. If the model isn't robust, explainability methods can't be trusted. In addition, the explanations need to be validated and robust enough to be trustworthy. One suggestion can be an ablation study to test the impact of boundary region variation.

Robustness cannot be estimated reliably. Due to the fact that the human monitoring is only conducted every 12 years, the system is currently trained on labels that are out-of-date while using recent satellite data as the input data. The robustness of the model will only be available some years in the future when the output data is compared to the manual inspection data (ground truth). Until new data is available, reliable estimation of the system's real performance and robustness is difficult. However, it is likely that manual visual inspection of satellite images would increase the validation capability to much larger areas meaning that ecologists can extend the mapping capability.

***Recommendation:*** The next manual mapping could be performed twice. Once an ecologist performs the mapping without knowledge of the output produced by the AI system and another ecologist uses the manual mapping to explicitly validate the values produced by the AI system both from the satellite imagery and manual inspection. This would allow for an estimate of how reliable the values produced by the AI system are and allow for a cross validation of human experts to detect, for example, subjective bias. Further, testing both the manual inspection from satellite data and manual inspections provide an increased understanding of human ability vs. model ability.

### 3.2.5. Explanations

The system uses an explanatory technique based on identifying individual pixels in the satellite imagery as grass or not. To some extent certain areas have been physically inspected to identify correctness of the classification. Given that each pixel represents an up-sampled 10m resolution provides the system with a fidelity that allows for detection of macro structures in the data.

**No interactive visualization.** The evaluated prototype is provided in a notebook form and no additional tooling has been provided to interact with the model output. While the developers are using a GIS based tool for inspection, this may demand a certain technical skill that not everyone possesses. To empower the product owners and their organizations in evaluating and interacting with the solution a more interactive visualization technique could be used.

**Visualizations do not provide information on variables.** While visualization is an important part for humans to understand the model output, the importance of understanding variable significance for the model output should be determined. For example, the importance of certain bands is not detailed for the different areas. While maybe this has not been detailed as a requirement for the prototype it would allow the developer to explain to ecologists more clearly what the variables represent and to detect potential confounding variables.

**Visualizations do not provide information on errors.** The solution improves upon the human error rate and achieves a score higher than demanded from the project owner. Still the solution is not perfect.

**Explanations of out-of-distribution areas.** The selected areas present a selection bias. The influence of human experts helps in the design of the system. Therefore, the continued use of the solution must involve the same selection process. The assumption must be, until proven otherwise, that the system does not generalize its findings to areas outside the scope of what has been tested. More extensive testing must be performed in order to understand the explanatory aspect of the model in areas outside the current ones. Topographical features of land surfaces could have some impacts on the data and the model performances. This aspect is not explored, features may also have influence on the generalizability of the models as well. Aggregated summaries on findings of grass growth should not be developed without the complete understanding of areas when the model is performing erroneously and why it is performing erroneously. Providing a context in relation to the explanation may help to understand such cases.

*Recommendation:* A more extensive evaluation of explanatory techniques is needed. A recommendation is that an evaluator tool is constructed with a low usability threshold, such as a web page. Said tool can also include the ability to inspect results visually, even down to a pixel level to understand variable importance for the outcome. Further aggregated metrics per area that measures variable importance should be developed to further understand the generalization power of the model. When more areas are tested these can then be compared and evidence can be presented that show the model's generalization power.

**Useful explanation technique for end-users is unknown.** The AI system makes its decision based on a time-series of values for the different spectral bands for a single pixel. While these explanations might adequately highlight the importance of the different bands for the system's decision, the usefulness of these explanations to human users is unclear.

***Recommendation:*** The usefulness of the explanations should be validated with interviews and studies. Ideally the explanations are shown to multiple representatives of the different user-groups that will likely use the system in the future to ensure that the explanations provided by the system are understandable and useful for each of these groups.

### 3.2.6. Deployment & Inference

The prototype evaluated has reached TRL6 level and therefore has not been deployed to a production environment. It has been tested though in terms of a number of pre-selected areas in a laboratory like environment. The setup of inference has been to understand the ability to classify samples that are in close proximity to the training samples. The development team has taken measures to avoid any mixing between training dataset and test dataset, and selected inference samples based on a certain distance to train samples. While the inference outcome has been deemed successful in terms of correctness metrics presented, visual interaction with the inference result has a certain skills threshold.

**Intended use is not completely clear.** Currently it is not completely clear how the model will be used once it reaches sufficient maturity. Two use-cases are possible. (1) The manual mapping is performed every 12 years, as currently is the case. The model will then be used on this data and current satellite imagery to refine the spatial resolution of the manual mapping and to interpolate the mapping to the years where no mapping data exists. (2) The manual mapping is performed more frequently, but only for a small part of the affected regions. This data is then used to train and update the model, which will then classify satellite imagery of other regions for which no mapping was performed in the current year. Depending on the final use-case the importance of different issues highlighted in this report will differ.

**Areas for inference overlap with training.** To improve the understanding of how well the model infer correct results, it should be tested on areas outside the training areas. Mixing areas for training and testing with an additional selection bias may create problems further down the line. To ensure that the model performs inference correctly the tested areas should not be part of the model training. Further, several tests should be performed to understand the importance of variables in each area and if the variable significance remains stable.

**Ambiguous cases in inference.** In creating the training data, the team allowed certain pixels to be classified as ambiguous and were consequently discarded. During inference this implies that results do not match real-world performance. The handling of hard to categorize data during inference needs further attention.

**No live feedback process.** A feedback mechanism as used in active learning systems could be useful in the long term to improve the system's accuracy. Thus, ecologists would need to be able to indicate/correct samples they do not agree with. Also during additional physical visits to areas under investigation a feedback process would allow the ecologists to validate model outcomes in a detailed fashion.

**Model card missing.** It is not immediately clear to users under which circumstances the model is valid and how it should be used. This poses a dependency on developers and a risk if employed by others. It should be clear to the user what the model does, what its intended use is and what potential pitfalls or biases there are. The developer should provide a short document with key information about the model for a broader audience, including for example managers, ecologists and IT staff. It should state the context in which the model may and, more importantly, may not be used. This document may be in the form of a model card.

### 3.2.7. Data Management and Processing

**Data stored in the company repository.** All data is first downloaded from a service and is then stored in a project repository. The data storage solution employed is sufficient for prototype testing. For system testing the product owner needs to document the handling of data in more detail and potentially provide an authenticated connector over a ciphered end-to-end connection to the data storage solution.

**Periodic model re-calibration required.** The developers indicated that, due to slight differences in time, the model needs to be recalibrated every year, otherwise the model's accuracy will decrease. It is not clear what knowledge and skills, data and technology are needed for this. There should be clear instructions for recalibrating the model, and a description of what kind of and how much labeled training data are required. Also it should be clear what the requirements are for the technical infrastructure.

**Machine Learning Operations (MLOps) process descriptions are missing.** To ensure reproducibility and eventually the monitoring of the continued use of the system, an MLOps process should be considered. Detailing and documenting an MLOps process will enable the retraining of the system and avoid vendor lock-in. Operations should without exception be managed through pipelines that detail each processing step to achieve the classification result.

### 3.2.8  Results of the mapping

The mapping to the EU trustworthy AI framework using a mapping from "open to closed vocabulary" as a consensus-based approach is presented in the Appendix.

## 3.3 Ecological Assessment

In addition to ethical and technical assessment, domain experts - in this case ecologists - assessed the AI system by evaluating the effectiveness of its performance in providing data that can be informative and unambiguous under the ecological perspective.

### 3.3.1 Assessment

The ecologists looked at the AI system in two ways.

**1. From manual to automated:** The transition from a completely manual field monitoring system, that is, carried out by ecologists on the field, to an automated system based on recognition of patterns in satellite imageries, has been assessed. The study in Friesland has moved from a fully human-controlled field-based land monitoring system with an accuracy of around 60-70% (as exposed verbally by the leaders of the field team section) into an automated Satellite Remote Sensing-based Artificial Intelligence-algorithm system with an accuracy of around 70-90% (as presented in the technical report ran by the company in charge of the AI-system) for mapping the selected two nitrogen-indicator species (*Molinia caerulea* and *Avenella flexuosa*).

**2. The algorithm:** Using the algorithm to create maps based on satellite imagery to monitor the natural area and the ecological processes that are undergoing. The ecologists raised three main concerns. These were then linked to the EU ethical principles and requirements for trustworthy AI.

Among other considerations, this report focuses on the Key Principal Indicators (KPI) and the 3 main concerns raised by the field experts in the Friesland province (see Appendix for more complete ecological discussions with the field experts in Friesland).

The concerns are (described in more detailed in the rest of this section):
1. The model tends to underfit: In some cases the model classifies less pipe straw compared to ground truth (estimated by experts).
2. The model classifies only 2 (unwanted) nitrogen-sensitive species, but there are more species that determine the healthiness of the heather fields.
3. Managers are reluctant to use this (or any) model if they are constantly seeking proof that the model is performing according to the desired KPI's. It's not just this classification model, but every model has trust issues.

KPI for the Natura2000 monitoring system: The model must have an accuracy of >80% in determining the presence or absence of both *Molinia caerulea* and *Avenella flexuosa* compared with the available data (latest vegetation maps).

**Concern 1**: Like many remote-sensing based systems, this AI system reduces costs, increases the frequency of monitoring (annual instead of every 12 years), and offers better spatial resolution (10mx10m pixels) than the manual polygons. Also, compared with multiple observers gathering data in different parts of the Natura2000 site, this AI-system offers higher monitoring consistency: whatever bias it has, it is likely to be consistent along the entire site. All these benefits are relevant for prompt action at ecological and policy levels. These benefits would be irrelevant, however, if the accuracy of the final maps was not equal to or better than the field-based monitoring (KPI). While this AI system remains under evaluation (e.g. temporal performance), it has already offered acceptable overall accuracy in comparison with the manual field-based monitoring. The system offers conservative results (does not include false-positives). Thus, some areas tend to show lower expansion of *grassification* (Friedrich et al., 2011) than in reality, but the system does not misclassify grasses that are not nitrogen indicators as false-positives, supporting its conservative responses. In other words, the AI-systems suffer more from omission errors (missing data) than from commission errors (false positives), which is a good response for conservativeness reasons.

Full accuracy is the desired outcome of any mapping, but when that is not the case (as it never is), conservativeness is an important principle for ecosystem management (less detected area under grassification than there is on the ground = "omission errors"). Conservativeness avoids unnecessary expenditure, field efforts, and wrong action directed to the wrong places (which would happen instead with "commission errors" in the satellite-derived map).

## 3.3.2 Benefits in the application of the AI system for the Province of Friesland

The use of artificial intelligence to monitor a vegetation community undoubtedly has significant advantages, in terms of time-saving, efficiency, and, possibly, consistent data accuracy and representativeness both in space and time. However, from a technical point of view, it is necessary to assess what are the limitations to the reliability of the AI algorithm. The limiting factors are usually related to the quality of the input data, e.g., the spatial and temporal resolution of the remotely sensed data and the presence of reflectance artifacts (the latter problem is particularly evident when dealing with wetlands and study areas that include water-rich soils, or with areas located in the nearing of water bodies).

**Ecological benefits:** The system captures well the two Nitrogen-indicator species under study and has the potential to move to multitemporal assessments, improving the current 10-year monitoring cycle. Considering the speedy advance of grassification (2- 4 times faster now than in the '50s), the need for prompt restoration action as well as prompt policy decision-making (limiting new licenses for certain activities, etc.) is clear. This AI system represents an improvement from the previous long-term manual approach (every 12 years). Simultaneously, better temporal and spatial resolutions on the location and advance of grassification will offer further clues about Nitrogen-attribution (does it relate to existing maps of nitrogen-atmospheric deposition? Or perhaps relates more to other sources of Nitrogen pollution, as to say fertilizer run-off, soil pollution, etc?).

**Policy implications:** Caution is suggested when connecting the final mapping with the legal consequences of heather degradation and economic activities in the surroundings. While the current mapping offers some good initial results, multitemporal data and better attribution of Nitrogen pollution would be required, even with field samplings, before moving toward legal action.

**Concern 2:** Some concerns have been raised by local experts managing the Nature 2000 area in Friesland, about the AI-system not classifying all the vegetation types that reflect nitrogen-driven degradation of the heather. The AI-system, however, has been trained to map only *Molinia caerulea* and *Avenella flexuosa*. Therefore these concerns, while very relevant, are out of the current AI system performance. Future monitoring should integrate other ecological needs such as changes in the extent of de-grassification and a more comprehensive assessment of heather responses over time (beyond nitrogen-pollution).

### 3.3.3 Issues and Risks

Open issues/claims where no evidence was found are potential risks that this AI system may pose, when it is used by the Province of Friesland and the ecologists working for them. When connecting the mapping to environmental or policy-making decisions, this AI System has an overall good performance (accuracy-wise) with a conservative response (trend to underestimate grassification but not to mix it with non-nitrogen-indicator grasses). These two responses make this system a relatively "low-risk" method for ecologically managing the two selected species (*Molinia caerulea* and *Avenella flexuosa)*.  However, the trend to under-estimate grassing effects has been raised as a concern by local ecologists working on the Nature 2000 area in Friesland (**concern 1**).

Under-estimation of grassing effects is also expected when the patches of *Molinia caerulea* and *Avenella flexuosa* cover an area smaller than 10x10 m$^2$, which is the current minimum mapping unit of the satellite data (Sentinel 2). If smaller patches than 10x10m are relevant for the goals of the Natura2000 site, the transition to an AI monitoring system will still require human support to track the presence of these smaller responses to nitrogen-pollution.

So far, the AI-System in the Province of Friesland has proven to be an efficient monitoring system for locating undesired patches of nitrogen-driven grasses in the heather, at a given moment in time. This is already very relevant for an effective, accurate, and efficient use of resources and human efforts (policy action for prompt restoration). However, equally relevant from the ecological perspective would be to understand grassification trends over time. This AI-system could potentially remain accurate in detecting the trends in grassification (and model future trends), but it has not yet been tested for multi-temporality. When multitemporal responses are assessed, and thresholds of grassification change defined, the system could potentially be an efficient early-warning system that raises attention to areas moving beyond certain thresholds of grassification, and an efficient performance monitoring system to assess restoration action.

**From an ecological perspective:** Local ecologists in the Friesland province have shown concern about the system only focusing on two species of nitrogen-pollution, while more species intervene in the health of the heather (**concern 2**). While the system has not been trained for that, this concern may need to be solved from both an ecological and an economic perspective.

The Key Principal Indicator for the Natura2000 monitoring system is the capability to have an accuracy of >80% in determining the presence or absence of both *Molinia caerulea* and *Avenella flexuosa*. The model does perform near to such an accuracy and captures well the two Nitrogen-indicator species

However, to analyze the health of the heather, not only grassification processes need to be tracked, but also de-grassification trends and stable-state conditions. The system is currently looking only into one environmental problem around nitrogen-pollution in heathers, rather than a more holistic management-oriented monitoring. More trends are needed to close the ecological cycle of heather effects by nitrogen-pollution.

More environmental needs than just nitrogen-pollution are covered under Nature 2000 monitoring. The system is performing reasonably well to track the selected nitrogen-indicator species. However, Natura2000 areas have many more variables to measure. It is unclear how accurate this AI system is to monitor more variables. The cost-effectiveness of this system against the 10 million Eur cost of running fieldwork, cannot otherwise be assessed.

**From a policy and an ecological management perspective**, a risk of the current system is its lack of multi-temporality and an associated accuracy assessment to see if new training data would be needed every year. Land managers need to understand the evolution of their interventions, and they need to track over time the responses of nitrogen-indicator species, beyond their initial detection. At this stage, we do not know if the AI-System will remain useful/stable/accurate for multitemporal mapping. The Cost-effectiveness of this AI-System will be affected when including the time element, therefore also affecting its utility as a support tool for long-term environmental and policy action.

The AI system will reduce but will not fully substitute ground data collection. First of all, the AI model is treasuring ecologists' data, and is built on them. More field data is expected and needed, as Key Indicators need field validation both on space and on time. Ground validation will inform on the accuracy of the system but will also support the monitoring of under-going restoration activities.

In addition, we are facing the threat that the indicator species map is seen as something that is not clearly related to the Nitrogen enrichment map. Consequently, this may raise trust issues. Indeed, the policy implications of the results from this AI-system still remain unclear. Caution is therefore suggested when connecting the final mapping with the legal consequences of heather degradation and economic activities in the surroundings.

**A question of trust (concern 3):** Ecology does not only dwell on and address environment-related questions but focuses also on the mutual interactions in the social-ecological system under study. In this case, the social-ecological system is represented by the Friesland ecosystem and human activities within it. We are living in an era where the separations between Nature, society, and science must be avoided: this is especially important when considering trust in ecology and the technological tools the ecologists use.

From a stakeholder's point of view, it could be worrying to be seen as one of the responsible actors causing grassification of the heathlands. Therefore, an issue derives from the eventual impression in stakeholders' opinion that the decisions influencing the economic activities are going to be based on a "completely aseptic" mathematical model, without any personal involvement of the local ecologists and decision-makers. Such an impression could represent a risky issue, that can even increase the perceived distance and distrust the stakeholders feel between themselves, the scientists, and the authorities.

Lack of trust in this AI-system is also a problem that may affect ecologists running fieldwork. This could relate to several factors such as lack of technical understanding, insufficient information, lack of communication, the black box nature of AI, etc.

However, empirical evidence of trustworthiness will come with accuracy assessments. While there often is a prior belief that human measurements are more reliable than automated algorithms, this belief requires accuracy assessments to be validated

**Results of the mapping**

The mapping to the EU trustworthy AI framework using a mapping from "open to closed vocabulary" as a consensus-based approach is presented in the Appendix.

### 3.3.4 Recommendations

This AI system performs better (improved accuracy than the previous human-made maps) for the selected task of identifying two Nitrogen-driven species (*Molinia caerulea* and *Avenella flexuosa*) in space, in the Province of Friesland. It, therefore, represents an accurate and conservative tool for land managers and politicians to track the location of patches of heath degradation and to act upon them promptly. Please note that human products are not necessarily better and should not be trusted better than other methods. Accuracy offers neutral empirical proof to define what is working best and, therefore, what should be trusted more (for the purpose of locating nitrogen-pollution problems). In our case study, when assessing the space dimension overall, the AI-system outperformed the accuracy of the human maps (as verbally defined to be ca. 60-70% of the real location of grassified patches). The next question would be to assess its performance on time. Based on our assessment, we make the following recommendations:

(1) Run accuracy assessments to validate its multi-temporality performance. Without the time component, it remains unclear of its full potential, benefits, and costs. Its functionality as an early warning system, and as a performant monitoring system on land interventions, depends on it.
(2) If new species are needed to offer a more coherent vision of Nitrogen-driven health degradation, new studies will be needed on the performance of this AI system. It is for the land managers to define this point, as much as which other ecosystem variables would be required to better comprehend heather dynamics (grassification and de-grassification).
(3) Legal consequences linked to the AI-system mapping would require further data on multitemporal responses of grassification and better attribution of the sources of Nitrogen pollution. Fertilizers (agriculture and/or livestock manure) dissolved in water run-off or aquifers, or stored in the soil, are most likely affecting these responses beyond nitrogen-atmospheric deposition).

In addition, we have the following general recommendations for the Natura2000 systems in the Netherlands and in Europe:

(1) An aim for **standardization**. New initiatives on remote sensing systems with or without AI algorithms are starting to appear in the Netherlands (*e.g. the Province of Utrecht is starting a pilot project to monitor vegetation changes in the Oostelijke Vechtplassen area, a [Natura2000 area](#) in Utrecht, with the use of remote sensing data).* The recommendation would be a dialogue among Nature2000 managers on their pilots, their needs, their indicators, and the performance of their tests, so that one common system may be applied to answer the same monitoring questions in the entire country.

(2) More clarity on the **overall monitoring needs for Natura2000 besides Nitrogen-indicator mapping**. This pilot seems an isolated exercise from the otherwise long list of variables/indicators needed to monitor and report under the Natura2000 network. A pilot that tested multiple variables to track multiple monitoring demands under the Natura2000 areas would offer a better understanding of the cost-effectiveness of remote sensing methodologies with or without AI algorithms. It is difficult to grasp if this AI-system is a good investment or not (e.g. the 10-million manual monitoring cost must have included many more variables).

(3) **Simplification:** Because of the multiple monitoring needs associated with integrated land management, simplification is key. One system that accurately responds to multiple indicators, would be preferred to one system that is very topic-specific. Automated monitoring systems need to respond to multiple demands. Cost-effectiveness will also need to integrate these needs.

(4) **Consistency**: Whatever method is selected now, should remain over time to avoid changes in statistics and trends related to the algorithm rather than to the land dynamics. Better to spend time trying different options than changing methods along the way.

(5) **Keeping some level of ground validation and fieldwork**: It is always advisable that the Natura2000 monitoring network/system retains some field data collection of key environmental parameters. If forced to consider just remote-sensing data, for example because the area is too large, it would be better to consider a system sensitive to more than one factor, as opposed to systems focused on only one. It would also be advisable to collect appropriate data to calibrate the satellite observations and run performance assessments of the quality of the maps over time. Fieldwork should not fully disappear.

(6) **Adding field data and ecological models results**: Working into an ecosystem always brings the necessity to have an holistic perspective and take into consideration not only the elements, but also interactions occurring in the system. More research into limiting factors in heather fields, as well as the factors driving the spatial and temporal distribution of N enrichment and grass patches, would improve the scientific base of the monitoring efforts, especially to keep track of future trends.

(7) **Provide participatory process opportunities for stakeholders:** to avoid the impression that decisions possibly ruling the stakeholders' activities are taken from a "completely aseptic" mathematical model, it is recommended to enhance stakeholders' involvement and provide timely communication to inform them about new data and updates of the system over the coming years. In addition, the effort to communicate the reliability and trustworthiness of the AI system will help raise awareness of the environmental problems associated with their production activities, so that they themselves can intimately understand how they are affecting the ecological processes of heatherland zones. Consequently, an opportunity is given to stimulate stakeholders to run self-motivated and self-regulatory interventions, or at least to better understand the importance of the reasons why regulatory interventions based on the results of the AI system are adopted.

(8) **Provide training and workshops for local scientists:** Lack of trust in this AI-system is also a problem that may affect ecologists running fieldwork, due mostly to lack of technical understanding or communication, insufficient information, etc. Therefore, it is advisable that data and information about the AI system are provided to local scientists, aiding to avoid the "Einstellung effect" (Luchins, 1942).

# 4. Ethics and Fundamental Rights Assessment

In light of the introduction of a fundamental rights impact assessment tool for algorithms (The Impact Assessment Fundamental Rights and Algorithms – [FRAIA](#)) in the Netherlands in March 2022, this Z-Inspection® process also included an assessment using parts of this tool. The FRAIA is envisaged as a discussion and decision-making tool for government organizations who commission the development and/or use of an algorithmic system. Like the Z-Inspection® process it aims to facilitate an interdisciplinary dialogue. See Section 5.1 describing the differences and similarities of the two approaches.

In identifying the fundamental rights being affected by the AI system, this assessment looked at the list of fundamental rights provided in the FRAIA. According to Annex 1 to the FRAIA document the rights are divided in four clusters of fundamental rights:
1. the person
2. freedom-related fundamental rights
3. equality rights
4. procedural fundamental rights

Specific rights are listed under each of the four areas. This assessment considered whether the rights under each cluster was affected by looking broadly at how such rights were set out in International or European Law (without doing any legal assessment), as well as how they are understood as ethical issues more broadly when considering the terminology used in the FRAIA. The first step was to consider how the AI system might support the advancement of a right, or affect it negatively, i.e., infringing on it.

In this pilot, for each of the rights identified as potentially affected, the assessment concludes with a *claim* whether the right is a) affected (regardless of whether this is positively or negatively affected), b) not affected, or c) might be affected depending on certain clarifications. A brief argument is made in respect of each claim and *evidence* is provided in support of whether the right is affected. The assessment identified five fundamental rights clusters which were potentially affected by the AI system.

**I. Rights related to the Person:**
1. Rights related to Healthy living Environment
2. Rights related to Personal identity/personality rights/personal autonomy
3. Rights related to Protection of data and informational privacy rights
4. Rights related to Territorial privacy

**II Procedural Rights**
5. Rights related to Right to good administration

Each cluster of fundamental rights is analyzed in the following by using the approach called *Claim, Arguments and Evidence* (Bloomfield & Netkachova, 2014; Brundage et al., 2020). Moreover, we will also present the results of analyzing the Ethical implications of using such an AI system as indicated by the EU Trustworthy AI framework (AI HLEG, 2019).

This approach is unique, as it combines a fundamental rights assessment with a Trustworthy AI Assessment using an evidence based approach.

# 4.1 Rights related to the Person

### 4.1.1 Right to a Healthy living Environment

In the FRAIA, healthy living environment rights in this cluster concern rights such as, right to sustainable development, right to environmental protection, protection from emissions of harmful substances and right to water (Gerards et al., 2022).
National courts have found that the right of a Healthy living Environment imposes the duty for the governments to take active and effective measures against climate change; if they fail to do so, the fundamental right to a healthy living environment might be infringed. This is evidenced by a Netherlands supreme court decision that the Dutch government is obliged by Art.2 ECHR (which protects the right to life), and Art.8 ECHR

(which protects the right to respect for private and family life), to reduce the greenhouse emission by at least 25 % as in order to protect the environment.

The court referred to the case-law of the ECHR (ECtHR) in which the Court stated, that these rights also include the obligation that the city must take measures if there is a real and imminent danger to the life or well-being of a person (*ECLI:NL:HR:2019:2007*, 2019).

Therefore, the government is obliged to take active measures in order to achieve the corresponding climate targets.

The AI System aims to enable a more frequent and precise, or unbiased, detection of the invasive and unwanted pipestraw in heatherfields than the manual classification by humans. The information from the AI System could be used to inform environmental policies, and potentially laws, which may impact livelihoods of individuals and groups in society, it could potentially also be used to inform administrative decisions in individual cases. The AI System is based on the assumption that the two species selected are useful proxies to determine reactive nitrogen pollution.

However, even if the outcome is incorrect in a way where the information provided by the AI System leads to more environmental protection, a challenge to the AI System could lead to lack of trust not only in the AI used, but in the policies or other actions. This in turn, could negatively affect the right to a Healthy living Environment.

Based on the findings of the technical working group regarding robustness and the questions raised by ecologists to the soundness of linking trends in grassland to nitrogen deposition, there are risks that the right to a healthy living environment will be negatively affected by the AI system. These risks should be mitigated for the AI system to have a positive impact on the right to a healthy living environment.

Firstly, the Technical working group raised concerns with different aspects of the model and made some recommendations for improvements. This will be important for the overall trustworthiness of the AI System and therefore whether it will affect the right to a healthy living environment negatively or positively.

Secondly, important information was provided both from the ecologists using the model and outlined in the final report from Ilinox (ilionx, 2021) indicating that prevalence of the two grasses and the link to nitrogen is difficult to establish. That the AI System does not determine nitrogen levels should be understood by users and guide how it can inform policy making about the environment and assessments of nitrogen.

Finally, like with other activities, AI Systems contribute to $CO_2$ emission. How much $CO_2$ an AI System produces depends on the way the AI is designed and used. This should be further considered and understood also for this AI System which aims to positively affect the environment.

The main aim of the AI system was to affect the healthy environment rights positively, however the assessment found that whether the rights are affected positively or negatively would depend on both the robustness and accuracy of the AI system, as well as on the negative impact $CO_2$ emissions from the AI system might have.

The assessment considered that the right to health living environment was engaged, due to the overall aim of the AI system and that by using a trustworthy AI System to assess the levels of nitrogen in the environment, the AI could provide sound information for responsive policy action with the potential to inform, and potentially advance the reduction of reactive nitrogen. However, should the AI system lack trustworthiness this could have negative and adverse consequences.

It was further found that the trustworthiness of the AI system will be affected by the accuracy, robustness, explainability and effectiveness of the AI system which in turn will be important factors in whether the system will affect the right to Healthy living Environment positively or negatively. This includes the validity of the foundational assumption that the prevalence of two types of pipestraw in heather fields is a valid and useful proxy for assessing nitrogen levels.

In case the outcome of the AI System is incorrect, in a way that the system is under-detecting the invasive and unwanted pipestraw and wavy hairy grass, it could result in insufficient measures to protect the environment, thereby negatively affecting the right to a healthy living environment. A system which is ineffective in this way could lead to a loss of vulnerable vegetation and disturbing biodiversity thus loss of complete ecosystems and therefore negatively infringe the right of Healthy living Environment.

**Ethical considerations - healthy living environment:** The above reflection is based on a fundamental rights perspective. From an ethics perspective, the central importance of environmental wellbeing, protecting the environment, and preserving biodiversity for individual and societal well-being and for future generations can only be further underlined. This is particularly true as the state and quality of the environment not only affects a very large number of humans, but also has wide implications on the wellbeing of a broad spectrum of species and biodiversity overall. Thus, the environmental wellbeing needs consideration not only from an anthropocentric perspective, but also from a pathocentric and biocentric perspective. While there do exist obligations related to Natura2000 and requirements that are based in current legal frameworks, there certainly is a general difficulty to balance the very broad concept of "environmental well being" against whatever step is taken in the context of the AI system. The preceding analysis is further complicated by the need to consider *future generations*. It is challenging to effectively bring in arguments that relate to the rights and requirements of future generations.

In the analysis of this use case, the question of who or which group works in favor of supporting environmental wellbeing and who or which group represents the interests of future generations is worth detailed consideration. One way of looking at this is assuming that the local authorities / the government assume this vicarious role to facilitate environmental well being and represent the interests of future generations. The open question at this point is whether local representatives / the government are effective in achieving these goals. While a positive outcome of the use of the AI system is expected, there are also potential negative outcomes with negative implications for biodiversity and the environment if the system's monitoring provides inadequate results.

**Mapping ethical issues to the seven requirements of the EU High Level Experts Group (HLEG)**

**Social and Environmental wellbeing:** The sixth out of the seven requirements the HLEG identified to be necessary to achieve trustworthy AI is "societal and environmental wellbeing" (AI HLEG, 2019, p. 14). It is divided into 'sustainable and environmentally friendly AI', 'Social impact', and 'Society and Democracy'.

The first of the three sub-requirements is to ensure sustainability and environmental responsibility of the AI-System, not only by ensuring that the AI-System is using resources in an environmentally friendly way, but also that research is advanced to use AI in global interest like the sustainable development goals. According to the HLEG AI Systems should ideally be used in a way it benefits citizens including future generations (AI HLEG, 2019, p. 19).

**Technical robustness and safety:** According to the aforementioned, the question whether the fundamental right of a healthy living environment is affected is directly linked to the effectiveness of the AI-System and therefore to "technical robustness and safety" as one of the seven requirements of the High Level Expert Group. Technical robustness and safety requires that the system is developed in a way harm is prevented (AI HLEG, 2019, p. 16). This necessarily requires that the AI-System is accurate in a way that the outcome of the system is based on proper classification and correct predictions.

The complete mapping to the EU trustworthy AI framework using a mapping from "open to closed vocabulary" as a consensus-based approach is presented in the Appendix.

***Recommendation*:** Implement the recommendations from the Technical Working Group to the AI System. The Technical Working group has pointed to some concrete steps required to improve the AI system and improve the robustness and accuracy. Without improving the robustness of the AI system the accuracy will be compromised. This in turn can lead to loss of trust in the system and the policies it informs. Reduced trust in environmental policies increases the risks of backlash which can have unintended negative outcomes for the environment.

*Recommendation:* Ensure that the limitations of the AI System in terms of predicting nitrogen levels are understood by users, including policy makers; and use this to inform what the AI System can be used for with integrity and trust (e.g. as an early warning system for further action to assess nitrogen levels, or more broadly to monitor the biodiversity of the heatherfield ecosystem and the state of a nature area and the presence of unwanted pipestraw).

*Recommendation:* Make efforts to ensure that the $CO_2$ emission from the AI system is estimated and understood to inform the contribution the system can make to the environment and to be able to respond to any concerns about this.

## 4.1.2 Rights related to Personal identity/personality rights/personal autonomy

Other rights in this cluster, considered in the assessment, related to rights to personal identity/personality rights and personal autonomy. The assessment found that such rights were not negatively affected by the AI system.

It was considered that although Art.8 ECHR sets out that the right of personal autonomy can be infringed in case of occupation-related disputes, it is necessary that the consequences of work-related measures concern private social life (European Court of Human Rights, 2022a), which we did not find would be the case for this AI system. Therefore, Art.8 ECHR is not affected in this case.

**Ethics:** From an ethical perspective, personal autonomy and freedom of decision-making are central concepts with questions including a broader spectrum of aspects, than legal compliance.

Individual autonomy could be indirectly influenced by the algorithm in the sense that the use of the AI System and its monitoring of the biodiversity of the heather field ecosystem somehow aims at informing policies, laws or actions with the goal of protecting or improving the environment. The algorithm is instrumental in achieving this goal.

For example, farmers, business owners and others could only be allowed to farm or run their business if they set up certain modifications, behave in certain ways, use certain fertilizers, or introduce other changes. This could negatively influence their individual autonomy and room for maneuver in that farmers would have to adjust their farmland-related decision-making to the advice or policy related to the outcome of the algorithm. Insofar, the system may require them to make decisions they would not do otherwise. If the system does not produce correct results, the freedom of decision-making and individual autonomy of farmers, landlords, business owners and others to decide their own behavior and the right to self-fulfillment may be inadequately curtailed by a policy based on these inadequate results.

In this, it has to be seen that if the information provided by the AI system leads to stricter regulation or is instrumental in enforcing regulation which infringes personal autonomy, it is not the AI system itself that is causing the infringement but the regulation. The AI only plays an indirect role. It is debatable whether this indirect role is enough for considering the rights as negatively affected by the AI.

In view of this it may be questioned whether a policy or law related to protecting the environment, such as for example limiting the use of fertilizers, really can be seen as an infringement on farmers' autonomy. It is surely not an infringement on their ability to make decisions related to how they live their lives and the personal choices they make as individuals. It only slightly modifies the context in which they make their decisions. However, depending on how the measurements made by the AI system are (or will be) used by policy makers, i.e,. in the province of Friesland and beyond, there may be relevant negative implications on individual decision-making in that the number of acceptable choices for farmers or other groups may be considerably reduced. Thus, it could be argued that in the drafting of any new regulation there will be a requirement for the legislators to consider potential negative effects on individual autonomy.

If there should be privacy-related risks or a risk that personally identifiable information may be involved, additional autonomy-related questions would touch on the question of whether or not the persons affected by the system have agreed to its use. This indirectly implies the need to provide adequate information to the population.

### 4.1.3 Protection of data/informational privacy rights

Other rights considered under this cluster were rights to protection of data and informational privacy rights. Although there are risks in general, it was noted that the AI System is not intended to process any personal data and although it may collect some data linked to personal information; with the data collected and stored the fundamental rights protecting data and privacy are not affected.

Consideration was given to article 8 ECHR which encompasses the right of self-determination, that also includes the right to privacy in regards to the processing of personal data (European Court of Human Rights, 2022a, p. 56).

To be classified as personal data it is not necessary that the data subject can be identified (*Guillot v. France*, 1996; *Mentzen v. Latvia*, 2004), it is rather sufficient that a person can indirectly be identified based on elements which can be used to derive personal information (European Court of Human Rights, 2022b).

Therefore, even if the system will not be used to identify any persons, what is relevant is whether it could be used to collect relevant personal information, like movement patterns or other personal information, such as the structure of a property (Skrabania, 2021).

Satellite photos in general have the potential to infringe privacy matters and images can be used to infer things about individuals in some circumstances, for instance, where individuals are linked to images via geographic locations. This means that in theory, if there is more of the invasive grasses detected in the areas next to single farmers, one could suspect that these farmers are creating more nitrogen emissions than others. This is less about the AI System processing personal or private data about farmers than it is about how the information is used and what conclusions and decisions it can support. As such it is advisable to ensure the output of the AI System is used responsibly and for its intended purpose only.

As for the risk of privacy issues with the AI Systems specifically, we note based on the findings of the technical working group, that: the lowest possible resolution is 10x10m; one single pixel in the satellite image corresponds to a patch of 10x10m; with this resolution structures below this resolution are not visible in the image, but might influence the color of the specific pixel. It is therefore found that this is not sufficient to violate privacy.

In addition, it was found that the process does not include collecting or storing of personal, or private data. Rather, the model uses publicly available Satellite images available on the European Space Agency repository. The company downloads pictures from this repository, pre-processes the images, runs the model on the pre-processed images and produces grassing maps based on the model outputs. This is done by providing geographic regions with values, which can be used in a GIS system as map overlays. As such, the system is not saving or storing images, or sensitive personal or private data.

It was noted, that the Data Protection Officer ("Functionaris Gegevenbescherming" in Dutch), did not consider that the processing from this AI system falls within the scope of Article 35 of the GDPR, requiring a Data Protection Impact Assessment when there is a "high risk" to the rights and freedoms of natural persons as a result of the processing. This would indicate a similar assessment as we have come to regarding the potential of infringing the right to privacy.

**Ethics:** Third requirement of the seven requirements of the ethics guidelines on trustworthy AI is "Privacy and data governance" (AI HLEG, 2019, p. 14). According to this principle the AI System "must guarantee privacy and data protection throughout a system's entire lifecycle." This belongs to information generated by the system regarding the behavior of individuals and their preferences. Further Data protocol has to be set up with provisions concerning who can assess the data under which circumstances (AI HLEG, 2019, p. 17).

Privacy-related questions from an ethics perspective: From what the group learned about the system, the risk that personally identifiable information is processed seems to be very low.

In view of this, it may be argued that unless proven otherwise, it seems that there is no infringement on privacy or no additional privacy infringement beyond data that is already publicly available. The satellite images being used, for example the structure of a property as seen on the satellite images, is already publicly available, either by accessing the publicly available satellite images or on Google maps. However, from a big data perspective, it may be argued that even if the data is already available and does not contain personally identifiable information, if combined with additional information, personally identifiable information may be generated.

### 4.1.4 Rights related to Territorial privacy

Finally, the territorial privacy rights (Könings et al., 2010) were looked at under this cluster. It was found that this is not affected.

Territorial privacy usually refers to a person's private and personal space, like once home and has traditionally been focused on surveillance or recording targeted for observation of individuals.

In the case law relating to the territorial privacy rights of ECHR camera surveillance is always referred to as target observation (*Antunes Rocha v. Portugal*, 2005; *P.G. and J.H. v. The United Kingdom*, 2001; *Vetter v. France*, 2005; *Wood v. The United Kingdom*, 2004).

While high-resolution data from satellites raise privacy concerns, as set out in 5.1.3, the assessment found that there is no surveillance of individuals foreseen, and no violation of personal territory or space. Therefore, we found that the AI system does not affect territorial privacy rights.

## 4.2 Procedural Rights

### 4.2.1 Rights related to Right to good administration

Under the Procedural Rights cluster, the assessment looked at the right to good administration, and found that this right is potentially affected by the AI system.

The effect the AI System can have on the right has to be seen in the context of its use and who will use it for what. If the environmental monitoring done by the AI system is used in decisions affecting individuals the rights to good administration may be negatively affected if the AI system is either not accurate, or the contribution or role of the AI system in the decision is not, or cannot, be explained.

Good administration has different definitions in different jurisdictions, but often includes principles such as the right to be heard, impartiality, fairness, consistency, transparency, due diligence, balancing interests, human-centered and reasoning (Diamandouros, 2007; European Union Agency for Fundamental Rights, 2007; The Parliamentary and Health Service Ombudsman, 2009).

There is evidence that a right to good administration exists at the EU level and that it covers the principles of transparency, reasoning and due diligence in decision-making. This is specifically highlighted in the FRAIA with reference to the codification of such principles in the EU Charter of Fundamental Rights. While the EU Charter Article 41 concerns decisions taken by the institutions of the EU, the inclusion in the FRAIA document is evidence of the relevance also in the Netherlands.

A review of the principles and legal requirements for good administration in the Netherlands may be useful to ensure both compliance and full integration of the ethical principles in the deployment and use of the AI system. Article 41 of the EU Charter includes the rights "of every person to have access to his or her file, while respecting the legitimate interests of confidentiality and of professional and business secrecy" and the "obligation of the administration to give reasons for its decisions".

The assessment found that it is unclear if, and how, the AI system may be used in decisions affecting individuals. While the main purpose of the AI system is described as informing general "nature policy and management", it has not been fully clarified whether foreseen or future use of the AI system could also be to inform decisions affecting individuals, such as for instance requests for expansion of activities which affects nitrogen levels including husbandry or farming.

If readings from the AI system are used to inform administrative decisions affecting an individual, or specific groups of individuals, good administration would require, among other, that the individual, or the group, is informed about the AI system's contribution in the decision and that the decision-maker is able to explain, not only how information from the AI system contributed to the decisions, but also how the AI system functions. It would follow that where a decision relies on information provided by the AI system, so in this case, information about the amount of unwanted grass vegetation in heather fields as a proxy for nitrogen levels in a given area, this must be transparent and the affected individual must be provided sufficient understanding about how the AI System works.

Arguably, the same principles and considerations apply to development of public policy and nature management where the AI System provides information to either. The negative effects of the AI system on good administration could occur if the AI System is biased, inaccurate or in other ways lacks trustworthiness, or if the decision-maker cannot explain the AI system and how it affected the decision(s) or policies. As such, the principles involved in the right to good administration have close links to the several of the seven requirements for trustworthy AI of the EU, or those found in the OECD AI principles aimed at informing how governments can shape a human-centric approach to trustworthy AI.

There is currently lack of evidence on the technical robustness of the AI system, and further clarification is needed on how the system will be used, including whether it will, or might in the future, be used to inform the public administration also in decisions affecting individuals directly.

There were some concerns raised by ecologists about using the AI system, which could indicate that those working with the system are either not fully able to explain how information from the AI system contributed to the decisions or how the AI system functions, or have concerns with both. There are also views from domain experts indicating that the maps, while very useful for managing heathland areas, cannot explain the links with nitrogen as grassification is not only due to nitrogen.

**Ethics:** Among the EU ethical requirements to consider in relation to good administration are the following:

(1) Sufficient **human agency and oversight**, requiring that the decision-maker understands the information from the AI system in order to use it and weigh it correctly in the decision. Good administration requires due diligence, which is not possible if the AI system is based on a "black box", but may also not be possible if the decision-maker is insufficiently trained or informed about how the AI system functions.
(2) The **technical robustness and safety** of the system, so the readings are accurate and reliable and can be reproduced. If the AI system lacks technical robustness it can lead to individuals losing rights or entitlements, contrary to good administration.

(3) **Transparency** both in relation to how the AI system was used in the decision and what the information from the AI system means. This is closely linked to the requirement of reasoned decisions in good administration, where the decisions are both explained and communicated in a manner the affected individual can understand.

In sum, the assessment suggests some steps to be taken to mitigate risks of affecting the right to public administration negatively.

Whether the right to good administration is affected, both for decision-making in public administration or for development of public policy, will depend on the outcome of the assessments regarding the technical robustness of the AI system as well as from the ecologists on whether the two types of grasses are a sufficiently established proxy for nitrogen levels.

In addition, some concerns will require ongoing monitoring and mitigating action when the AI system is deployed and in use. This could include ongoing training of decision-makers, monitoring of the continued technical robustness of the AI system as well as the experience of using it in decision or policy making, as well as sufficient sector expertise, biology, within the administration to understand the information provided by the AI system.

*Recommendation*: In addition to improving the robustness of the system to the outcome can better inform the monitoring and management of heatherfields, more should be done to understand the reluctance by ecologists to work with the system.

*Recommendation*: To increase trust, explanations about the AI system and knowledge sharing on this should be available for those who will use it to inform policy as well as the public in general. This should be open source and available for public consultation and scrutiny. Ongoing evaluations of the model should be ensured, and the public administration should ensure sufficient resources, both technical and domain specific are available.

*Recommendation:* With the current limitations the AI System should not be used to inform decisions affecting how individuals are treated, such as in assessing requests for permits for economic activities or other.

# 4.3 Additional relevant aspects that arise from the Trustworthy AI Assessment

We present here additional relevant aspects that arise from the Trustworthy AI Assessment that have not been considered in the fundamental rights-based type of assessment. The above part of this report describes the assessment results based on the FRAIA document and the fundamental rights-based approach, complemented by additional ethical considerations on the fundamental rights under discussion, i.e.

individual autonomy, privacy, territorial privacy rights, equality before the law, right to good government.

For the ethics assessment that follows the Trustworthy AI assessment process based on the Trustworthy AI guidelines, additional aspects related to ethical pillars, ethical requirements and sub requirements need consideration. For these, the Ethics Guidelines for Trustworthy AI, the broader context of relevant socio-technical scenarios around the use case, as well as the ALTAI questions, are relevant.

## 4.4 Ethical issues identified

In addition to the ethically relevant aspects discussed above in the context of the fundamental rights-based assessment, the following ethical issues were identified for reflection.

**Transparency and lack of transparency:** How transparent is the algorithm and the entire AI-based process and are the decisions based on the AI system well-founded? The system functions as a black box, it is difficult if not impossible to explain decisions made based on the results of the systems. It is currently unclear how well policies based on the system would be justified. One negative effect is that the decisions based on the system are not / would not be transparent and it would be difficult for the administration to explain how the system came up with its output and how well-founded the output is. Is the use of a non-transparent AI appropriate for a government, as the government cannot explain the system's results or decisions to the citizens? What will be the role of lack of transparency when promoting an unpopular policy? This is not only about the right to good administration but more broadly related to the Trustworth AI requirements of "Transparency" which includes as sub-requirements: traceability, explainability and communication and to the Requirement of Societal and Environmental Wellbeing. Questions to be asked include whether there is a third party (maybe us?) that will monitor the use of the AI system.

**Receiving relevant information:** From an ethics perspective, it could be argued that the fact that additional information about the environment is collected and provided to the wider public speaks strongly in favor of using the AI system. The system allows us to collect information more frequently. It provides information about the wellbeing of the environment more frequently than would otherwise be possible - even though it has to be stressed that the quality and reliability of the provided information is not clear at this point.

**Human agency and oversight:** Are decision-makers capable of explaining how the AI system came to its decision/results? Is it adequate to base relevant public policy decisions on a system we don't know how it came up with its results? In view of the lack of transparency, it is unclear whether the use of the AI positively or negatively influences the quality of the administrative process. Could using the AI system over a longer period of time involve de-skilling the ecologists currently performing the ecological surveys? How is this related to the goal of human-centric approaches to trustworthy AI? Is there a plan for continuous evaluation once the system is fully used? Is there a third party that monitors the use of the AI system (such as Z-Inspection®)?

**Technical robustness and safety:** How accurate is the system? Are those who are involved in the process and who may potentially make decisions based on the AI system aware that the results of the AI system are not always correct? Has adequate training been provided for people to better understand this? Are there concerns about cybersecurity? If so, are there potential ways to mitigate risks?

**Justice and fairness:** From an ethics perspective, potential policy implications of the AI system raise several questions with regard to justice and fairness. Several groups of the population could lose income and room for maneuver, whereas others could profit financially from the algorithm use. When or if the local authorities / the government considers policy-making based on the AI system, the following questions need to be addressed:

(1) Would it be fair to ask farmers to modify their farming, given that part of the excessive nitrogen does not result from their individual farming practices but from the broader context? This is probably a very general question that plays a role in a lot of current and similar contexts. Compensation payments?
(2) What if the measures of the AI system are used to support policy decisions that are potentially unfair with respect to certain groups of the population? E.g. Why target farmers and not other classes of population, for example car drivers or owners?
(3) What would be the role of the technical system in the larger political scheme? Would it serve to further support unpopular political decisions, deviating public attention from human decision-makers who are behind?

**Cost reduction:** Conflict of interest, in that some of the actors involved would lose income, whereas others would profit financially from the algorithm use. A technical system could be used as an additional argument in favor of implementing an unpopular political agenda.

**Diversity and Inclusion:** Who are the drivers behind the system? Is it in the interest of the entire population? Is there a plan to systematically include feedback from the wider population once the system output is widely available? Also, the needs and interests of the current population (including farmers) and future generations have to be balanced.

**Responsibility and Accountability:** It is unclear who is responsible for the model itself. Is the use of the system part of a uniform method of monitoring, or is this something the Province of Friesland chooses to use as an additional monitoring tool? Also, it is unclear who is responsible for the outcome of the algorithm.

**Due diligence in decision-making:** Can due diligence be granted if the decision is based on a "black box" AI? In order to avoid arbitrariness, the government would have to know for certain that the AI system produces adequate results. To what extent would "good administration" presuppose that the persons in charge understand the decision-making process? What extent of transparency, explainability and explicability would be required from the system?

How far does the outsourcing of relevant functions to private companies negatively influence the extent of control the administration/government has over the process?

What could be the societal implications of this lack of accountability, especially in a tense social crisis involving a major shift in social life, as can be seen from farmer protests.

# 5. Comparing the Trustworthy AI assessment process with the fundamental rights-based FRAIA assessment tool

In the pilot project in Friesland, the Z-Inspection® use case concerned an AI system to be used by a commune in the Netherlands. In March 2022, the Dutch Ministry of Interior and Kingdom Relations issued an Impact Assessment tool for Fundamental rights and algorithms (FRAIA) as a "discussion and decision- making tool for government organizations". The tool, introduced in section 1.1, sets out the questions which must be answered when a government organization considers "developing, delegating the development of, buying, adjusting and/or using an algorithm." The tool considers three decision- making stages for an AI system and asks that in all stages, respect for fundamental rights must be ensured.

For the Z-Inspection® use case in Friesland part of this tool and framework was included on a pilot basis. The FRAIA is a tool, which in many ways is at the forefront when it comes to ensuring that government organizations live up to human rights obligations and commitments. It foresees a multidisciplinary and holistic approach.

The working group responsible for the ethical and fundamental rights part of the assessment used two different assessment approaches: the fundamental rights-based assessment outlined by the document "Fundamental Rights and Algorithms Impact

Assessment (FRAIA)" (Gerards et al., 2022), and the Z-Inspection® Trustworthy AI assessment (Vetter et al., 2023; Zicari et al., 2021) based on the European Ethics Guidelines for Trustworthy AI (AI HLEG, 2019).

The FRAIA document suggests a procedure for assessing AI tools from a fundamental rights perspective that identifies whether an AI tool affects fundamental rights and, if so, facilitates a structured discussion about opportunities to prevent or mitigate this interference. The document defines four main clusters of fundamental rights: 1) fundamental rights relating to the person, 2) freedom-related fundamental rights, 3) equality rights, and 4) procedural fundamental rights.

## 5.1 FRAIA and Z-Inspection®: Similarities and Differences

The FRAIA comprises four parts of which the fourth part is on fundamental rights assessment. This assessment has a roadmap comprising seven steps. These steps can be succinctly explained as follows:
1. **Fundamental right:** does the algorithm affect (or threaten to affect) a fundamental right?
2. **Specific legislation**: does specific legislation apply with respect to the fundamental right that needs to be considered?
3. **Defining seriousness:** how seriously is this fundamental right infringed?
4. **Objectives**: what social, political, or administrative objectives are aimed at by using the algorithm?
5. **Suitability**: is using this specific algorithm a suitable tool to achieve these objectives?
6. **Necessity and subsidiarity**: is using this specific algorithm necessary to achieve this objective, and are there no other or mitigating measures available to do so?
7. **Balancing and proportionality**: at the end of the day, are the objectives sufficiently weighty to justify affecting fundamental rights?

There are some overlaps and complementary steps between the Z-Inspection® and the FRAIA. Where the Z-Inspection® is a *"process based on applied ethics to assess if an AI system is trustworthy, as set out by the high-level European Commission's expert group on AI",* the *"FRAIA aims at preventing the premature use of an algorithm that has not been properly assessed in terms of the consequences, entailing risks such as inaccuracy, ineffectiveness, or violation of fundamental rights, through asking a number of questions".*

To avoid duplications with other parts of the Z-Inspection® , only the questions in step 1 and 3 of the FRAIA roadmap for the fundamental rights assessment have been included in the pilot, while the other questions of the fundamental rights assessment were considered in the broader context of the Z-Inspection®.

As indicated by (Ulrich, 2023) "a potential downside of the FRAIA approach is that it may be perceived as excessively demanding and cumbersome by practitioners not specifically trained in the area of human rights and for whom this field of inquiry remains secondary to the objectives driving the original engagement with AI. Experiences shared by some Z-Inspection® focal groups echo this concern. An important forward-looking challenge will therefore be to devise relatively simple and intuitive, yet comprehensive templates for human rights impact assessment."

In contrast, the Ethics Guidelines for Trustworthy AI are based on four mid-level principles: Respect for Human Autonomy; Prevention of Harm; Fairness; and Explicability.

The guidelines describe seven key requirements closely connected to these ethical principles: Human agency and oversight; Technical robustness and safety; Privacy and data governance; Transparency; Diversity, non-discrimination, and fairness; Societal and environmental well-being; and Accountability. Broadly speaking, the FRAIA assessment relies on fundamental rights, whereas the Trustworthy AI assessment is based on ethical principles.

When working on the two assessments, the following three questions arose:
1. **Human Rights - framing:** How to frame the human rights assessment? Should this be considered differently from an ethical assessment, or the assessment of trustworthiness?
2. **Assessment - the aims:** What should be the aim or scope of the fundamental rights assessment? Is it to ensure that rights are not infringed or violated only, or does it go beyond to look at how rights are affected more broadly, included, protected or promoted?
3. **Trust and decisions about the AI system:** Some ethical issues impacting trustworthiness of the AI system concern the decisions about using the AI system more broadly, beyond a rights assessment.

While the FRAIA tool was useful and clear, the question about how to frame the human rights assessment nevertheless arose and more specifically, how to consider the fundamental rights as part of an assessment of trustworthiness and ethical reflections on an AI system. Should we consider the rights as they are defined in law and interpreted through the courts only? Or should the rights be considered more broadly, as part of the assessment, linking the rights to ethical principles beyond their narrower legal definition? If only the legal definitions are used, an assessment of whether specific legislation applies would be required.

This may on the one hand have the advantage of ensuring adherence to existing human rights definition and legislation, while on the other hand be narrow in scope, and therefore risk missing broader ethical questions. It might also require specialized legal expertise and could risk excluding other perspectives. In the use case, we adopted a hybrid approach, by first identifying which rights in the framework could be affected, using the legal definitions as part of the argument of why the rights were engaged; before turning to the broader ethical issues related to the rights. In this way, a more open reflection was possible, integrating the fundamental rights framework as part of a broader ethical assessment. This approach was used as the Z-Inspection® is not aimed at assessing legal compliance.

If the human rights assessment is defined too narrowly it risks being an assessment separate from the ethical assessment, or the assessment of trustworthiness. If it is too broad, the human rights standards risk being watered down. A two-tiered, integrated approach, looking both at legal requirements and the broader ethical questions, could be envisaged, depending on the organizational set up and use case.

## 5.2 The FRAIA Assessment

### 5.2.1 Considering if rights are infringed, protected or promoted?

Frequently raised issues regarding AI and human rights concern how personal data is handled and used in the AI system, right to privacy; or how the data might be biased and can lead to discrimination. The Dutch use case did not concern personal data. The FRAIA suggests that the right to equal treatment, protection of personal data, procedural and good administration rights should always be considered, as these are usually affected by an AI system. However, fundamental rights may also be infringed or affected by the implementation, use, or application of the algorithm, by the context in which the algorithm is used, or by the decisions and measures that are linked to the output of the algorithm. This was highlighted in the FRAIA and is at the center of the Z-Inspection® process which assesses trustworthiness based on the socio-technical scenario.

The FRAIA considers as a first step the identification of fundamental rights which may be affected, or threatened, by the AI system and then a balancing of the seriousness of the rights infringement with the importance of the objectives and the necessity to use the AI to reach the objective. The FRAIA also includes a framework for assessing seriousness. In the practical application of the FRAIA in the Z-Inspection® use case two adaptations were made. Firstly, it was decided to also include, in the assessment, the rights which the AI system aimed to affect positively, i.e., the right to a healthy living environment. It was found that the objective of the AI system was largely to promote this fundamental right and that it was useful to include this perspective in the assessment. The objective of promoting a right was considered *"a claim"* in the assessment, rather than a fact, and as such arguments and evidence for the claim were discussed.

Secondly, the FRAIA sets out a four-tier framework for assessing seriousness of the rights infringement. This was complex to use and was replaced, in the use-case, by considerations and suggestions for how risks of any infringements could be mitigated. Taking this broader approach to consider if rights were potentially infringed, or sought to be protected or promoted by the AI system was found useful.

### 5.2.2 Ethical issues arising from the decisions about using an AI system.

The FRAIA asks that the questions about "why an AI system" and "what the system should do" are answered before the question on "how will the AI will do this" is answered. The "why" questions in the tool are questions like "What are the reasons, the underlying motives, and the intended effects of the use of the algorithm? What are the underlying values that steer the algorithm's deployment?"

The tools provided, in support to answer this, are directly linked to guidance and the parameters for sound policy and law making (Dutch Ministry of Justice and Security & Center for International Legal Cooperation, 2017). Central to answering this are questions like "who are involved in the problem formulation and the articulation of the solution?" In other words, the ability for an organization, or government, to design, develop, deploy and use trustworthy and human rights aligned AI will be directly affected by its overall ability to answer such questions as part of governance, accountability and pursuit of legitimate aims. This in turn is directly linked to how power is distributed and managed both within the organization and with its stakeholders and collaborators. As the experience from the Z-Inspection® use-cases shows, this requires not only clarity in the problem formulation, the objective and goal statements, and in the reasons given for why an algorithm is most useful, but also a process to answer this. What came out in the assessment is that, to generate trust, such a process must reflect different viewpoints, ensure accountability and be legitimate and transparent to its stakeholders.

Answering questions about why an AI solution is considered, what it intends to do and what the underlying values are can be challenging to do in practice and answers to such questions may differ across the organization. As such it may be unclear how this will be ensured and how this will be ensured, and what is expected of an AI system developer who is asked to find a solution to a problem which lacks these qualities and legitimacy. However, in the context of the Z-Inspection® process the importance of clarity in the problem statement and objectives came to the fore, as did the anthropocentric approach implied in human rights. Experiences from the use-cases indicate that for AI systems to be trustworthy it is important to have
   1. Clarity in the problem formulation - "what problem do *we* want the AI system to help solve".

2. Have shared values and ethics reflected in the "why" this is a problem and the suggested solution.
3. Legitimacy of the "we". In other words – "do those who are deciding and were consulted on the problem and solution have legitimacy in the eyes of those affected by the decisions both in the short and longer term? And how are they held to account?"

One aspect of this is ensuring clarity on how ethical issues are identified and dilemmas solved. It follows that where decision-making processes or organizational governance are weak, the decisions related to the design, development, deployment and use of an AI system risk undermining trust or posing ethical questions beyond human rights assessment.

## 5.3. Lessons Learned from the two assessment approaches

The fundamental rights assessment and the ethics assessment based on the Trustworthy AI guidelines go hand in hand; both approaches provide critical insights with regard to the AI use case. Reflecting on AI from an ethics perspective clearly overlaps with a fundamental rights assessment. Both ethics and fundamental rights are about norms and fundamental values held in society. As ethics reflection and ethics guidelines influence law, scholars from both fields must work together when thinking about the shaping of technology and its societal implications. Even though there are great similarities, there are several considerable differences between the two approaches.

Ethics, a branch of philosophy, reflects on what is right and wrong. It seeks to find an answer to questions like "What are we to do?" or "What is the right action?". In the context of AI applications, an ethics-based approach addresses questions like "What is the right way to design, develop, deploy, and use this type of technology so that it is beneficial for individuals and society"? Questions like these require thinking about the various alternatives for action around an AI application, and involve reflecting on the various options and their potential implications without confining the reflection to those options in line with existing law.

A fundamental rights-based approach is more closely linked to existing law and focuses on aspects that are legally relevant and thus enforceable. Compared to this, an ethics-based approach is much broader and also more open to reflection on potential implications that may not be worth considering from a legal perspective. For example, from an ethics perspective, personal autonomy, freedom of decision-making, and fairness were found to be concepts of clear relevance in the context of the pilot project's AI tool, whereas, from a rights-based perspective, rights related to personal autonomy in a strictly legal sense were considered not infringed by the AI tool.

While a fundamental rights-based assessment focuses on whether fundamental rights are negatively affected or infringed, from an ethics perspective, both positive and negative implications of AI technology are considered. In this pilot, for example, the potential positive implications of the AI tool on the environment proved to be central. This implies the question of whether the right to a healthy living environment may or may not be positively affected by the AI tool. Furthermore, a fundamental rights-based approach towards protecting the environment is clearly anthropocentric, as can be seen from the wording "right to a healthy environment". The FRAIA document lists the right to a healthy living environment in the cluster "Rights related to the person".

In contrast, an ethics-based perspective allows us to bring in biocentric or pathocentric perspectives and address biodiversity-related issues. Whereas anthropocentric positions focus on human beings when it comes to moral and legal considerations, pathocentric ethical positions consider (human and non-human) suffering as morally relevant, whereas biocentric positions ascribe value to all forms of life. Accordingly, these latter ethical positions allow us to consider issues related to climate change, the environment, and biodiversity more directly. Also, from an ethics perspective, questions of how to adequately consider future generations can be tackled more easily than from a fundamental rights-based approach. Overall, the fundamental rights-based approach clearly funnels and constrains the aspects, questions, and issues to be discussed around the AI use case. For example, issues related to transparency or human agency and oversight can only be addressed in the context of the right to good administration, even though transparency and human agency and oversight are clearly relevant in other contexts as well. As discussed above, similar problems arise in the context of the right to a healthy living environment. Approaching the use case from a fundamental rights perspective implies that ethical and societal aspects and implications of AI are discussed only in-sofar as they are related to fundamental rights and existing law.

In conclusion to this section, we quote (Ulrich, 2023) who suggested four levels of consideration that should be taken into account. These are:

"1. *Possible direct adverse human rights impacts*; this, in fact, is the exclusive focus of the FRAIA assessment tool.

2. *Possible indirect adverse impacts in the form*, e.g., of reinforcement of existing inequalities, patterns of structural discrimination, and further marginalization of disadvantaged social groups, etc., due to algorithmic biases and to a gradual remodeling of work, employment, and social and economic access.

3. *Capacity of AI to facilitate and positively contribute to the (progressive) realization of human rights* (e.g., in the health sector) and other related societal objectives as, e.g., defined by the Sustainable Development Goals.

*4. Fundamental challenges posed by AI to some of the core underlying premises of normative reasoning* such as, notably, the concepts of human dignity, agency and autonomy, free will, intentionality and accountability."

# 6. Discussions

In this section some key aspects of the experience of having worked in this pilot are presented.

## 6.1 Choice of experts

This is a voluntary, non-binding assessment of potential ethical and human rights concerns and issues that may surface in the use of the AI system. It is designed to complement, rather than supplant, other assessments focused on the AI system's compliance with relevant laws, standards, and regulations.

The choice of experts required for the use-case has an ethical implication since the quality of the analysis and the results depend on the diligent selection and quality of experts. This includes the experts not being biased or in a position of conflict of interest (Vetter et al., 2023).

In this use case, the design, and implementation of the AI system were outsourced to a third-party vendor. We adopted the policies that the third-party vendor would not be part of the assessment team to avoid any conflict of interest, and the main use-case owners would need to declare that they do not have any involvement with the third-party vendor that could lead to a conflict of interest (Vetter et al., 2023).

The third party vendor shared useful technical information that was evaluated independently by the technical working group. The third-party vendor was not involved in the assessment nor in providing recommendations.

## 6.2 Impact

At the time of writing this report, the AI System is not used by policy makers at the Province of Friesland. Partly due to the nature of this innovation project, plus the aforementioned concerns resulting from this assessment. The Province of Friesland is currently implementing the suggested improvements that are thanks to the outcome of the assessment in this pilot project.

The Z-Inspection® results have proven to be a valuable contribution towards the aim of having a human-centered approach to using AI responsibly. This is not only regarding the improvement of the technical aspects but also the relevance of broader implications.

The following two subsections are a description of the main lessons learned when conducting this pilot project by the Province of Friesland (5.2.1 Technical recommendations and 5.2.2. Lessons learned.)

### 6.2.1. Technical recommendations

Regarding the technical aspect, there were several recommendations that have been addressed and implemented to improve the pilot model. The recommendations were regarding the training, execution and validation of the model. The recommendations were sorted by their impact/complexity in order to plan their implementation accordingly. Some of the recommendations that have been implemented are:

1. A model card to illustrate transparently how (well) the method works
2. Clearly defined the metrics by which the results are assessed
3. Quantitatively reporting on how well the methodology extrapolates spatially, (i.e. how well does the model score within versus outside of the training dataset)
4. Quantitatively reporting on how well the methodology extrapolates temporally, (i.e. how well the model performs on years within versus years outside of the training dataset)

### 6.2.2. Lessons learned

**Technical and Ethical validation of AI systems is a requirement.** It is apparent that validations of AI systems should include both technical and ethical assessments. A broader perspective is needed when assessing AI systems.

**Create a common language, clearly defined scope and roles.** It is important to establish a common language within a newly formed group, this requires time, understanding and patience. It is also important to be aware of the diverse backgrounds, especially regarding the significance of context within language. It helps to differentiate between professional and cultural aspects when discussing a topic. The spoken/written language, especially at the start, caused some miscommunications and/or misunderstandings between each other. Throughout the assessment, it is crucial to establish a well-defined common language, clearly defined scope and appoint a leading role who consistently monitors and proactively helps to establish a common language. The Z-Inspection method helped with this as it clearly defined common assessment aspects and process steps.

**Increase awareness across the organization.** The Z-Inspection® results have proven to be a valuable contribution towards the aim of having a human-centered approach to using AI responsibly by including ethical, fundamental and ecological assessments. This is not only regarding the improvement of the technical aspects but also the relevance of broader implications within the government. Firstly the method facilitates awareness that civil servants need to understand the impact AI can have on their work processes and society.

**Create a selection matrix.** Another crucial discussion point revolves around the necessary capacity and time investment associated with the Z-Inspection®. Therefore the relationship between optimized results, the AI model's complexity and the required investment needs to be taken into account when selecting an assessment methodology. For future projects formulating a set of guidelines as a selection method when to choose the Z-Inspection® or when to select other evaluation methodologies would be beneficial. The more complex high-risk cases would benefit more from a full Z-Inspection® whereas others the Fundamental Rights and Algorithms Impact Assessment [FRAIA] would be enough or a combination of the two.

In conclusion, the Z-Inspection® methodology clearly illustrated the benefits of evaluating the pilot project broader by including ethical, fundamental and ecological assessments. Both from a technical perspective as well as from considering if rights are infringed, protected or even promoted. This was an intensive and time consuming inspection that required working proactively and transparently which resulted in useful feedback and improvements. Even though as of this moment there are no policy implications on an individual level using the results from this AI-system, the experience proved to be valuable and raised awareness for both policy makers, technological team and strategic management.

## 6.3 Novelty

We applied a novel approach to the assessment of fundamental human rights for AI which is based on *supporting verifiable claims* (Bloomfield & Netkachova, 2014; Brundage et al., 2020). This is an integral part of the approach when using the Z-Inspection® process.

Integrating The Fundamental Rights and Algorithm Impact Assessment (FRAIA) into the Z-Inspection® method contributed to great conversations about human rights, both in the pilot and during the Z-Inspection® Venice conference (Z-Inspection® Initiative, 2023).

The FRAIA is a comprehensive questionnaire. The FRAIA barely answers the question of what a trustworthy process for conducting a human rights assessment should meet. For example, what agreements should you make in advance to avoid conflicts of interest? How do you ensure that it is clear from the outset with whom the results will be shared, and how do you clarify intellectual property? How do you organize the assessment of ethical, legal and technical aspects of an AI system in relation to each other, without sacrificing depth and rigor? How do you assess whether claims are properly substantiated? The Z-Inspection® method is a valuable complement to FRAIA because it addresses these and other questions and provides a framework for orchestrating human rights reviews based on FRAIA.

Introducing the concept of technical readiness levels also enriched FRAIA. It offers the possibility to take into account the stage of the life cycle the AI system is in when answering the questions in the FRAIA and put more or less focus on certain questions.

Applying the Z-Inspection® method to the case study also resulted in a set of recommendations to take the model to the next technical readiness level. This is also complementary to performing a FRAIA. The FRAIA focuses primarily on identifying human rights risks, much less on collecting recommendations to improve the system.

Taking into account the technical readiness level also means you can draw a more nuanced conclusion regarding the system and make targeted recommendations to take the system to the next level. It is not the case that a system is either compliant or not. The question is what level the system is at and what it will take to get to the next level. That's a lot more positively worded and can therefore gain more support.

## 6.4 Technology Readiness Levels (TRLs)

Performing an assessment of an AI system before its technical conception or during its early maturity state, as in this pilot, demands a thorough Z-Inspection® setup phase to determine the in-scope requirements for the assessment. Design time terms such as XX by Design, with XX representing aspects such as privacy, security, and/or accessibility, usually assume that the system will end up in real-world use.

However, the experience from several technical TAI assessments using Z-Inspection® has shown us that this is not always the case. AI, at least in its current state, demands plenty of experimentation, likely more so than non-AI solutions. Hence, assessing such solutions requires us to contemplate more about the current Technology Readiness Level as well as the planned TRL the project intends to reach.

There are several considerations related to the TRL level to take into account in the setup stage before the actual discovery process of claims begins. First, we can consider the level of technical details from the ideation stage. Depending on the case owner's background and the amount of preparation time, some may have detailed diagrams, notebooks, and/or component specifications. Others may have more normative descriptions of intended possible outcomes. While all documentation is useful, the challenge can arise with more normative descriptions that continue to evolve and essentially lead to a cat-and-mouse game between the assessors and the development team. Every time the assessors question a certain aspect of the idea, the development team uses this as input to refine their system. Therefore, in cases where the assessment is performed at a very low TRL state, a time constraint should be considered both on how long the assessment can take and when the development team is allowed to redefine their product. Otherwise, the challenge becomes one where no one understands the technical meaning of the assessment outcome. If we may generalize, we can say that the TAI assessment timeframe should not exceed workshops over 1-2 days during the ideation stage of a new solution. Furthermore, a recommendation is to avoid the use of normative language that is too unspecific. Terminology such as robust or transparent are problematic in technical translation as they do not reveal enough information about the relative aspects in the meaning of the terms. Z-Inspection® experience shows that having transparent AI is very difficult to achieve in practice (Amann et al., 2022) given the complexity of models and data, while also the technical understanding of this term can be considered illusive, given that we are unable to assert specific metrics for such non-functional requirements.

Second, during the inception stage, when software architects start to formulate concrete functional and non-functional requirements for the AI solution, there is more to build an assessment upon, and it also becomes a fruitful discussion to challenge perceptions and assumptions going into the design. During the setup phase, such clarifications can already reveal insights and help projects focus on the core design. Furthermore, logical flaws during the inception stage can become very costly for the project down the road, and discovering these early by enforcing sufficient documentation or explicit presentation can help establish a more expressive product management style. Additionally, during the inception stage, the assessment team must start establishing clear definitions for the TAI requirements of the project. Here, we have found tools such as the web based prototype tool Assessment List for Trustworthy Artificial Intelligence (ALTAI) (AI HLEG, 2020) supportive but not quite sufficient in real-life assessments. A static tool such as ALTAI lacks the adaptations needed for the system context. The context of the assessment is always important, and the discovery of technical issues depends on a thorough description of the stakeholders, their intentions, users, and potentially different user stories, as well as the environment where the system will be used.

Third, progressing to the development stage of a project provides a new level of detail such as code modules, data understanding, preliminary modeling results and targets, and pipeline automation metrics. Depending on the TRL the project aims for, these will become more and more detailed. Additional maturity targets are introduced as non-functional requirements (such as security or privacy). However, at this stage, it is also common to discover new non-functional requirements from other expertise areas that were not considered or assumed as facts during the technical part of the assessment. This can relate to domain expertise failing to interpret results or how to consider accessibility demands. This is where static tools or checklists often fail to support a good assessment. We have found that sometimes the newly discovered issues demand either a refocusing of the project or TAI assessment scope or that a redesign is really necessary in order to provide more clarity to what the developer's real claims are. As a project matures through the development phase, which spans several TRL levels, its TAI requirements naturally increase with maturity.

Here, it is also important to understand the interlinkage of organizational maturity to handle AI systems in production as well as the system under review. An AI system does not exist in isolation, and in particular, those based on complex deep learning models demand a constant stream of data to work and to be monitored for potential drift and bias issues that can emerge at any given time. In an early development phase, the organization must start to prepare for a potential release of the system to a test group.

Fourth, during the system test phases, going from data-at-rest validation of modeling to system acceptance testing in a real organization, the organizational maturity must develop an understanding of how to continuously maintain such systems according to a set of developed requirements and metrics. From our assessment experience, many organizations that have not developed sufficient maturity for AI operations and ethics miss this crucial part. The focus on AI as a technology often overshadows the operationalization of AI in the organization. Here, technical methods and tools such as DataOps and MLOps can address parts of the operationalization process, but AI operational maturity demands a culture that embraces the fact that, at one point, the AI system will fail. Working out fallback and feedback mechanisms and human oversight is crucial for an organization that embraces the fact that the AI system will fail at some point in the future. Through extensive monitoring and adversarial testing, some potential risks with AI can be mitigated, but for safety-critical systems, we should never assume that they are infallible.

## 6.5 Transparency

This report is made publically available. The results of this pilot are of great importance for the Dutch government, serving as a best practice with which public administrators can get started, and incorporate ethical and human rights values when considering the use of an AI system and/or algorithms. It also sends a strong message to encourage public administrators to make the results of AI assessments like this one, transparent and available to the public.

## 6.6 Limitations and Challenges

As described in (Vetter et al., 2023) one inherent limitation of this process is that its success depends on good-faith cooperation from the use-case owners that go "beyond compliance". For this assessment, we relied on the use-case owner to provide us with the relevant information, and to implement our recommendations for mitigating the discovered ethical issues.

The assessment team consisted of experts that volunteered their time and received no compensation for their work in the assessment. An independent third-party assessment is highly desirable, however this also imposes limits regarding the invested time and can lead to uneven contributions in both quality and quantity.

Communication - especially language - was a barrier. The assessment was conducted in English. Although all members in the pilot spoke good English, some members were not used to discussing some specific aspects in English, especially the ethical and legal aspects. This sometimes made it difficult to respond sufficiently quickly and comprehensively in a discussion.

In addition, scientists and policymakers each have their own jargon. It took time to understand each other. This highlights the importance of a common language to achieve a quality assessment of the system.

# 7. Conclusions

The Province of Fryslân has developed a new monitoring system prototype based on remote sensing. The pilot project analyzed whether this AI system is trustworthy and can be used responsibly in practice. The experience, results and lessons learned in conducting a pilot project "Responsible use of AI " - a cooperation with the Province of Friesland, Rijks ICT Gilde- part of the Ministry of the Interior and Kingdom Relations (BZK) (both in The Netherlands) and a group of members of the Z-Inspection® Initiative- were presented in this report. A novel approach to the assessment of fundamental human rights and Trustworthiness for AI where the Fundamental Rights and Algorithm Impact Assessment (FRAIA) was integrated into the Z-Inspection® process was presented. Lessons learned will be fed into the further development of FRAIA within the Dutch government.

At the beginning of our assessment, we considered the prototype as a moderate technical innovation (relatively well-known technology) and anticipated a straightforward integration process into an existing organizational workflow. However, through the technical part of the assessment and workshops with developers and product owners, we identified several areas in extension of the core innovation requiring improvements and validations before the system can be effectively adopted for the intended organizational use. Remote monitoring via satellite imagery holds significant promise for governmental environmental monitoring efforts. Yet, a classification solution, in isolation, merely provides data about the analyzed pixels. When we aggregate classification results into area summaries, descriptive statistics, or decision-making outcomes, it is crucial to approach the extension of this solution into a 'system of systems' with precision and specificity. Although the prototype outperformed human experts in some instances, it was not without flaws. Reliance on probabilistic decision support means that errors, when they occur, not only propagate through a system of systems but also magnify (compounded errors). Unlike a chain of human experts, who can identify and correct logical errors through understanding of the decision-making context, a machine learning toolchain lacks the ability to detect such logical inconsistencies, presenting a markedly different challenge. At the end of the technical part of the assessment, this area where organizational processes meet technical solutions, provided many interesting and insightful discussions with the whole team.

## Acknowledgments

We would like to thank Rosa Maria Roman-Cuesta who was instrumental in the ecological part of the assessment.

# Affiliations

**Marjolein Boonstra**, *Strategy Consultant & Programme Manager; Rijks ICT Gilde, Ministry of the Interior and Kingdom Relations, The Netherlands.*

**Frédérick Bruneault**, *Professor, École des médias, Université du Québec à Montréal and Philosophie, Collège André-Laurendeau, Canada*

**Subrata Chakraborty**, *School of Science & Technology, University of New England, NSW, Australia.*

**Tjitske Faber**, *Team manager i-Strategy and Innovation, Province of Friesland, The Netherlands*

**Alessio Gallucci**, *Z-Inspection® Initiative, Utrecht, The Netherlands.*

**Gerard Kema**, *Innovation manager, Province of Friesland, The Netherlands*

**Heejin Kim**, *Visiting Assistant Professor, Graduate School of Data Science, Seoul National University, Seoul, Korea*

**Jaap Kooiker**, *Strategy Consultant; Rijks ICT Gilde, Ministry of the Interior and Kingdom Relations, The Netherlands.*

**Eleanore Hickman**, *Lecturer in Corporate Law and Governance, University of Bristol School of Law,, Co-director of the Centre for Private and Commercial Law and member of the UK Young Academy*

**Elisabeth Hildt**, *Professor, Illinois Institute of Technology, USA; Fellow at L3S Research Center Hannover, Germany*

**Annegret Lamadé,** *PhD Candidate, Institute for Law and Regulation of Digitalisation, Philipps-University Marburg, Germany, and law clerk at the Hanseatic Higher Regional Court, Germany*

**Emilie Wiinblad Mathez**, *Z-Inspection® Initiative, Geneva, Switzerland*

**Florian Möslein**, *Professor of Law at the Philipps-University Marburg, Germany; Director of the Institute for Law and Regulation of Digitalisation; and Vice-Director of the Centre Responsible Digitality,Germany*

**Genien Pathuis,** *Innovation managerNetherlands*

**Giovanni Sartor**, *Professor, Cirsfid-Alma AI and Department of Law, University of Bologna, Italy; and European University Institute of Florence, Italy*

**Marijke ter Steege**, *Strategy Consultant; Rijks ICT Gilde, Ministry of the Interior and Kingdom Relations, The Netherlands.*

**Alice Stocco**, *Postdoctoral researcher, Environmental Sciences, Informatics and Statistics Dept., Ca' Foscari University, Venezia, Italy*

**Willy Tadema**, *AI Ethics Lead; Rijks ICT Gilde, Ministry of the Interior and Kingdom Relations, The Netherlands.*

**Jarno Tuimala**, *Independent researcher, Finland*

**Isabel van Vledder**, *Data Scientist; Rijks ICT Gilde, Ministry of the Interior and Kingdom Relations, The Netherlands.*

**Dennis Vetter**, *PhD Researcher, Computational Vision and Artificial Intelligence Lab, Goethe University Frankfurt, Germany*

**Jana Vetter**, *PhD Candidate; Marine Holobiomics Lab, Justus Liebig University Giessen, Germany*

**Magnus Westerlund**, *Principal Lecturer, ARCADA University of Applied Sciences, Finland; Associate Professor Kristiania University College, Oslo, Norway.*

**Roberto V. Zicari**, *Z-Inspection® Initiative Lead; Affiliated Professor ARCADA University of Applied Science, Finland; Adjunct Professor Graduate School of Data Science, Seoul National University, South Korea.*

# Grants

Giovanni Sartor was supported by the Grant: ERC-Advanced CompuLaw (grant agreement No 833647)

Dennis Vetter received funding from the European Union's Horizon 2020 research and innovation program under grant agreement no. 101016233 (PERISCOPE), and from the European Union's Connecting Europe Facility program under grant agreement no. INEA/CEF/ICT/A2020/2276680 (xAIM).

Roberto V. Zicari was supported by the Brain Pool Program through the National Research Foundation of Korea (NRF) funded by the Ministry of Science and ICT ( grant number: 2022H1D3A2A01082266, Research Title: Assessing Trustworthy AI).

# APPENDIX

## Mappings

This appendix lists the mapping to the EU trustworthy AI framework. Each working group used a mapping from "open to closed vocabulary" as a consensus-based approach.

### Technical Working Group

| |
|---|
| **ID Ethical Issue: E1,** Unclear use-case of the model and how to put it into practice |
| **Description:** Currently, the model is only available in the form of a jupyter notebook, as well as outputs in the form of a GIS map. As is, it provides a nice proof of concept that the general task of estimating the grassing from satellite images is technically feasible. For a future fully integrated and deployed system, it is lacking a clear definition of final use-case, as well as capabilities to efficiently reproduce and monitor the model. |
| **Map to Ethical Pillars / Requirements / Sub-requirements (closed vocabulary):** <br><br> Prevention of harm > Technical Robustness and Safety > Reliability <br><br> Explicability > Transparency > Communication |
| **Narrative Response:** Two use-cases are possible. (1) The manual mapping is performed every 10-12 years, as currently is the case. The model will then be used on this data and current satellite imagery to refine the spatial resolution of the manual mapping and to interpolate the mapping to the years where no mapping data exists. (2) The manual mapping is performed more frequently, but only for a small part of the affected regions. This data is then used to train and update the model, which will then classify satellite imagery of other regions for which no mapping was performed in the current year. <br><br> Depending on the final use-case the importance of different issues highlighted in this report will differ. The importance of E1, the model's ability to generalize, is especially affected by this: in scenario (1) generalization to other geographic areas not in the training data is much less required than in scenario (2), where this generalization is a must for reliable model outputs. |

In addition, detailing and documenting an MLOps process will enable the retraining of the system and avoid vendor lock-in. Operations should without exception be managed through pipelines that detail each processing step to achieve the classification result, as currently almost all knowledge is with the external company that built the system, which can lead to maintenance problems in future years.

**ID Ethical Issue: E2,** Lack of "gold standard labels" for training and evaluation

**Description:** The different ecologists who perform the mapping use different scales to describe the grassing of areas, which makes it unclear what labels are best and how they should be included in training and evaluation.

**Map to Ethical Pillars / Requirements / Sub-requirements (closed vocabulary):**

Prevention of harm > Technical Robustness and Safety > Reliability

Respect for human autonomy > Human agency and oversight > Human oversight

**Narrative Response:** For training, the model unifies these different scales provided by the ecologists into an ordinal variable with levels that do not necessarily completely overlap with the scales used originally. This can lead to problems in training where the system is not trained on the correct behavior. It also makes evaluation of the model and comparison between model and human experts very difficult. This is highlighted by the evaluation process, where agreements between model and human on less than 50% grassing or more than 50% grassing are considered a correct prediction.

**ID Ethical Issue: E3,** Model not proven to generalize to areas not in the training data

**Description:** The model was not shown to reliably generalize its performance to geographic areas not in the training data. This lack of generalization is supported by the need for yearly model retraining with updated satellite images.

**Map to Ethical Pillars / Requirements / Sub-requirements (closed vocabulary):**

Prevention of harm > Technical Robustness and Safety > Reliability

Explicability > Transparency > Explainability

Explicability > Transparency > Communication

**Narrative Response:** The model's lack of generalizability is supported by the need for yearly model retraining with updated satellite images for continually high performance. More effort should be put into an analysis of the model's failure modes and how reliable its performance is when used on new data. The limitations of the model should also be openly communicated with the stakeholders, as well as included in the model's visualizations to avoid a false sense of accuracy in end-users.

## Ecologist Working Group

**ID Ethical Issue: E1** The model tends to underestimate grassing effects

**Description:** The internal performance of the AI-system leads to lower detection of patches of *Molinia caerulea* and *Avenella flexuosa* than in reality. Moreover, due to the spatial resolution of the satellite data (10x10 m2), grassing effects happening at smaller scales than 10x10 m2 may remain undetected.

**Map to Ethical Pillars / Requirements / Sub-requirements (closed vocabulary):**

Prevention of harm > Technical robustness and safety > Reliability

Prevention of harm > Technical robustness and safety > Accuracy

**Narrative response:** the algorithms may need more training areas that cover more habitat situations to reduce omission errors. While less grassing effects is an error and it has ecological and policy implications, it is a better error than over-estimation of grassing effects (which we do not see in this AI-System)

| |
|---|
| **ID Ethical Issue: E2,** the System currently tracks only two nitrogen-pollution indicator species |
| **Description:** The AI-System has been trained to monitor *Molinia caerulea* and *Avenella flexuosa*. However, other less common nitrogen-sensitive species also exist within the protected area. |
| **Map to Ethical Pillars / Requirements / Sub-requirements (closed vocabulary):** <br><br> Prevention of harm > privacy and data governance > quality and integrity of data |
| **Narrative response:** the leaders of the pilot decided to focus on the most common nitrogen sensitive indicator species for this initial exercise. Some concerns have been raised on the tracking of the other nitrogen-sensitive species in the heather. |
| **ID Ethical Issue: E3,** the System currently tracks only nitrogen-related environmental problems |
| **Description:** Natura2000 areas have many more variables to measure. It is unclear how accurate this AI system is to monitor these other variables. |
| **Map to Ethical Pillars / Requirements / Sub-requirements (closed vocabulary):** <br><br> Prevention of harm > Technical robustness and safety > Reliability <br><br> Prevention of harm > privacy and data governance > quality and integrity of data |

**Narrative response:** Without evaluation of the need for the other variables it is not clear how useful the final system will be in replacing the manual mapping performed by ecologists

---

**ID Ethical Issue: E4** the System has trust issues among implementers

**Description:** People are being reluctant to use this model when proofs of model performance -according to desired Key Principal Indicators (KPIs)- are constantly being sought.

**Map to Ethical Pillars / Requirements / Sub-requirements (closed vocabulary):**

Explicability > Transparency > Communication

Prevention of harm > Technical robustness and safety > Reliability

**Narrative response:** It is recognized by local implementers that trust issues affect any classification model. This is not a specific problem of this AI-System, but trust issues are more relevant when moving from human monitoring to an automated system that relies both on satellite data and on AI algorithms. There is a sense of black-box effect where the classification process and the final map output remain poorly understood by map users. This is, however, a common problem with satellite-based classification mapping. Artificial intelligence adds a new layer of complication to the final output. The final goal of having more frequent equally accurate data and a less expensive monitoring system, needs to be contrasted with the loss of human control in the classification process.

---

**ID Ethical Issue: E5** Unclear how the results of the algorithm and its mapping will be used in the decision-making context

**Description:** The maps resulted from nitrogen-indicators species could be used as a tool to implement governance actions, raising concerns on the fairness about the translation of the results of the AI in the decision-making process

| **Map to Ethical Pillars / Requirements / Sub-requirements (closed vocabulary):** |
|---|
| Fairness > Accountability > Auditability |
| Fairness > Societal and environmental well-being > social impact |
| Fairness > Diversity, non-discrimination and fairness > stakeholder participation |
| Explicability > Transparency > Communication |
| **Narrative response:** fairness, consistency, and transparency with respect to the stakeholders who are seen as responsible for N excess in the heatherlands must be checked, leading to a participative process for a better understanding of the results. |

## Ethics and Fundamental Rights Working Group

If or when we begin with the fundamental rights for the mapping, there is a funneling effect in the sense that those aspects identified to fall under fundamental rights will be those that are considered "ethical pillars". This could be seen as an advantage as it avoids coming up with a broad variety of different aspects as starting points for the mapping, but it could also be criticized as an approach biased by the fundamental rights assessment/approach.

| **ID Ethical Issue: E1,** Healthy living environment |
|---|
| **Description:** There is a need to protect the environment and to secure sustainable development. |
| **Map to Ethical Pillars / Requirements / Sub-requirements (closed vocabulary):** |
| Prevention of harm > Technical Robustness and Safety > Accuracy |
| Prevention of harm > Technical Robustness and Safety > Reliability |

**ID Ethical Issue: E2,** Healthy living environment for current and future generations

**Description:** There is a need to protect the environment to avoid/prevent negative effects on human autonomy for current and future generations.

**Map to Ethical Pillars / Requirements / Sub-requirements (closed vocabulary):**

Respect for Human Autonomy > Diversity, Non-Discrimination and Fairness > Stakeholder Participation

---

**ID Ethical Issue: E3,** System may negatively affect individual autonomy

**Description:** Potential negative implications on human autonomy, self-fulfillment and individual decision-making for members of certain groups (farmers etc.)

**Map to Ethical Pillars / Requirements / Sub-requirements (closed vocabulary):**

Respect for Human Autonomy > Diversity, Non-Discrimination and Fairness > Avoidance of Unfair Bias

Respect for Human Autonomy > Diversity, Non-Discrimination and Fairness > Stakeholder Participation

---

**ID Ethical Issue: E4,** Respect for human Autonomy and information requirements

**Description:** Need to inform the actors and groups of actors involved or potentially affected by the system and to directly or indirectly obtain their informed consent.

| **Map to Ethical Pillars / Requirements / Sub-requirements (closed vocabulary):** |
|---|
| Respect for Human Autonomy > Transparency > Communication |

| **ID Ethical Issue: E5,** Is there privacy-related harm? |
|---|
| **Description:** Are there privacy-related infringements? |
| **Map to Ethical Pillars / Requirements / Sub-requirements (closed vocabulary):** |
| Prevention of Harm > Privacy and Data Governance > Privacy |

| **ID Ethical Issue: E6,** Would a good government use a system that is not explainable? |
|---|
| **Description:** Is it adequate for a government to use a non-transparent system the results of which cannot be explained? |
| **Map to Ethical Pillars / Requirements / Sub-requirements (closed vocabulary):** |
| Explicability > Human Agency and Oversight > Human Agency and Autonomy |
| Explicability > Human Agency and Oversight > Human Oversight |

**ID Ethical Issue: E7,** What is the broader influence of lack of transparency and explicability?

**Description:** Is it adequate to use a system that lacks transparency and explicability in a context with broad societal implications?

**Map to Ethical Pillars / Requirements / Sub-requirements (closed vocabulary):**

Explicability > Societal and Environmental Well-Being > Environmental Well-Being

Explicability > Societal and Environmental Well-Being > Impact on Work and Skills

Explicability > Societal and Environmental Well-Being > Impact on Society at Large

---

**ID Ethical Issue: E8,** Would policy based on the AI system be fair?

**Description:** Policy based on the AI system may have unfair implications

**Map to Ethical Pillars / Requirements / Sub-requirements (closed vocabulary):**

Respect for Human Autonomy > Diversity, Non-Discrimination and Fairness > Avoidance of Unfair Bias

Respect for Human Autonomy > Diversity, Non-Discrimination and Fairness > Stakeholder Participation

---

**ID Ethical Issue: E9,** Who is accountable for the system, its results and its policy implications?

**Description:** Who is accountable for the system, the results of the system and the policy implications?

> **Map to Ethical Pillars / Requirements / Sub-requirements (closed vocabulary):**
>
> Explicability > Accountability > Risk Management
>
> Explicability > Accountability > Auditability

**Identify tensions**

*Claim:*

There is a tension between the right of a healthy living environment and the personal autonomy of the farmers to be free in their decision-making.

*Arguments:*

In case the government makes a regulation based on the results of the algorithm to protect the healthy living environment, the government will, on the one hand, promote the fundamental right to a healthy living environment for future generations, but on the other hand, restrict the personal autonomy of the farmers on whom the regulations impose obligations.

This tension will have to be resolved by the governments when they are making the rules, that will affect farmers. The government therefore has to establish a proportionate balance between the fundamental rights by ensuring that the interference with personal autonomy is proportionate to the benefits that the regulation brings to the environment. Whether the fundamental right to a healthy living environment will justify the infringement on personal autonomy depends largely on the individual regulation made by the government.

Therefore, if governments are designing regulations, they must always balance those two aspects. In order to achieve a balanced relationship it could therefore help to repeatedly ask the question: how much does this regulation interfere with the personal autonomy of farmers? and relate this to the question: how much does regulation actually benefit the environment?